\documentclass[final,5p,times,twocolumn,authoryear]{elsarticle}
\pdfoutput = 1

\usepackage{amssymb}

\usepackage{svg}
\usepackage{amsmath}  
\usepackage{tabularx}
\usepackage{array,makecell}

\usepackage[flushleft]{threeparttable} 
\usepackage{booktabs,caption}
\usepackage{booktabs}
\usepackage{graphicx}
\usepackage{rotating} 

\newcolumntype{C}[1]{>{\centering}p{#1}}
\usepackage{microtype}
\usepackage[font=small,skip=0pt]{caption}
\usepackage{mathtools}
\usepackage{blkarray, bigstrut}
\usepackage{gauss}
\usepackage{multirow}
\usepackage{longtable} 
\usepackage{enumitem} 
\usepackage{xcolor,colortbl}

\usepackage[flushleft]{threeparttable}

\journal{Journal of Network and Computer Applications}

\begin{document}

\begin{frontmatter}



\title{NLP Methods in Host-based Intrusion Detection Systems: A Systematic Review and Future Directions}



\author[]{Zarrin Tasnim Sworna\fnref{l1}} 
\author[]{Zahra Mousavi\fnref{l2}} 
\author[]{Muhammad Ali Babar\fnref{l1}} 
\fntext[l1]{School of Computer Science, University of Adelaide and Cyber Security Cooperative Research Centre, Australia.}
\fntext[l2]{School of Computer Science, University of Adelaide and CSIRO/Data61, Australia.}



\begin{abstract}
Host-based Intrusion Detection System (HIDS) is an effective last line of defense for defending against cyber security attacks after perimeter defenses (e.g., Network-based Intrusion Detection System and Firewall) have failed or been bypassed. HIDS is widely adopted in the industry as HIDS is ranked among the top two most used security tools by Security Operation Centers (SOC) of organizations. 
Although effective and efficient HIDS is highly desirable for industrial organizations, the evolution of increasingly complex attack patterns causes several challenges resulting in performance degradation of HIDS (e.g., high false alert rate creating alert fatigue for SOC staff). Since Natural Language Processing (NLP) methods are better suited for identifying complex attack patterns, an increasing number of HIDS are leveraging the advances in NLP that have shown effective and efficient performance in precisely detecting low footprint, zero-day attacks and predicting an attacker’s next steps. This active research trend of using NLP in HIDS demands a synthesized and comprehensive body of knowledge of NLP-based HIDS. Thus, we conducted a systematic review of the literature on the end-to-end pipeline of the use of NLP in HIDS development. For the end-to-end NLP-based HIDS development pipeline, we identify, taxonomically categorize and systematically compare the state-of-the-art of NLP methods usage in HIDS, attacks detected by these NLP methods, datasets and evaluation metrics which are used to evaluate the NLP-based HIDS. We highlight the relevant prevalent practices, considerations, advantages and limitations to support the HIDS developers. We also outline the future research directions for the NLP-based HIDS development.


\end{abstract}




\end{frontmatter}


\section{Introduction}
Cyber security attacks increased by 1,885\% worldwide in 2021 \citep{SonicWall}. The attacks fueled by the profit-making  cyber-crime economy with nation-state targets are severely affecting every industry including healthcare, financial services and government organizations; costing 4.24 million dollars per attack breach on average globally in 2021 \citep{ibm_21}. To protect an organization by minimizing cyber attacks and thwarting new threats targeting an organization's IT infrastructure, a Host-based Intrusion Detection System (HIDS) is an effective last line of defense to detect and prevent malicious activities after perimeter defenses (e.g., Network-based Intrusion Detection System (NIDS), Firewall) have failed or been bypassed \citep{review_ids_20}. The use of HIDS is a prevalent security practice as HIDS consolidated its position as a preferred defense technique and ranked among the top 2 most used security tools by organizations among various commercial tools \citep{soar_rep_2021, insider_threat_2018}. HIDS helps organizations to identify malicious activities by monitoring host data (e.g., system call sequence) which, if left undetected, can lead to detrimental breaches. Even if HIDS could identify and prevent 5\% of all cyber attacks on time worldwide, it could help avoid a loss of around 260 billion USD \citep{s9}. Hence, performance enhancement of HIDS is an active research area due to the ubiquitous adoption of HIDS \citep{review_ids_4}. 

Despite all the research and industrial efforts, the evolution of the increasingly complex attack vectors causes HIDS to suffer from a high false alarm rate. The high false alarm rate of HIDS significantly contributes to alarm fatigue of the security team \citep{s4}. Other key challenges for the existing HIDS include handling a massive amount of system call traces, run-time attack detection by continuous monitoring of host data, keeping good detection rate, handling imbalanced data and reducing the required processing time and resource consumption. These challenges hinder large-scale HIDS deployment in commercial settings \citep{s41}.

To mitigate the above-mentioned HIDS challenges due to the evolution of the increasingly complex attack patterns, traditional HIDS techniques are becoming obsolete. Hence, researchers are introducing new Natural Language Processing (NLP)-based HIDS techniques which are better suited to complex attack patterns \citep{s43, s42}. HIDS researchers are motivated to benefit from methods that have been proven successful in the NLP domain, due to the high similarity between natural languages and System Call (syscall) sequences of a host used by HIDS \citep{s43, s44}. Syscalls denote the interface of user space programs for requesting services (e.g., read a file) from the Operating System's (OS) kernel. The way humans communicate with each other through natural language constructed based on specific grammar and syntax, host processes gain their intentions with syscall sequences, which are used by processes to communicate with OS in a host \citep{s44}. For HIDS development, a comprehensive and detailed source of understanding a host's behavior is syscalls executed in that host \citep{s9, s21}. This understanding enables fast detection of anomalies at run-time by analyzing streams of syscalls \citep{s21}. 

\begin{table}[h]
\centering
\caption{Example syscall sequences for reading two files \citep{s9}}
\label{tab:sys_ex}
\begin{tabular}{cc}
\hline
Seq1 & ``open, read, write, open, read, write”\\
Seq2 & ``open, open, read, write, read, write”\\
Seq3 & ``open, open, read, read, write, write”\\\hline
\end{tabular}
\end{table}
Consider a program that opens two files to read at the same time. When the program is running, the syscalls (e.g., open) sequences created by threads of the program may vary as shown in Table \ref{tab:sys_ex}. If the only sequence in the normal pattern library of a traditional HIDS is Seq1 and if the input sequence of HIDS for detection is Seq2, a traditional HIDS based on pattern matching may create a false alarm considering Seq2 as abnormal \citep{s9}. Hence, researchers consider syscall sequences of a syscall trace as sentences of a document with each syscall as a word, where the order, frequency and semantics of each word contributes to the understanding of host behavior \citep{s5, s46}. Since NLP has different methods for analyzing the context and semantics of a word, HIDS researchers are using NLP methods to analyze the contextual syscall similarity \citep{s54}, behavioral syscall semantics \citep{s57}, to preserve the sequence information of syscall sequence \citep{s54} and to learn correlation of syscalls \citep{s5}. 

Researchers are developing NLP-inspired problem formulation for HIDS development such as the use of question-answer models for generating future syscalls considering previously invoked syscall sequence as question and generated syscall sequence as answer \citep{s63} and considering host intrusion detection as classifying a sentence into normal or abnormal that is similar to sentiment classification \citep{s46}. The latest HIDS research trend is the adoption of NLP methods considering from simple NLP methods such as the use of n-gram and Term Frequency-Inverse Document Frequency (TF-IDF) to extract contextual and statistical features to advanced NLP methods such as word embedding, semantic ontology and language modeling for domain adaptation, knowledge fusion and next syscalls prediction. The use of NLP enabled the HIDS researchers to develop HIDS with more accurate intrusion detection capability of unknown attacks in real-time by processing syscall streams on the fly \citep{s5, s30} and the ability to predict a future syscall sequence possibly to be executed during an attack \citep{s63}.

Despite the drastically growing adoption of NLP in HIDS development, there has been relatively little effort allocated to systematically analyze and synthesize the available peer review literature to understand how NLP is used in HIDS development. The lack of a synthesized and comprehensive body of knowledge on such an important topic motivated us to conduct a Systematic Literature Review (SLR) of the papers on the end-to-end pipeline of the use of NLP in HIDS development. For the end-to-end NLP-based HIDS development pipeline, we identify, taxonomically categorize and systematically compare the state-of-the-art of NLP methods usage in HIDS, attacks detected by these NLP methods, datasets and metrics which are used to evaluate the NLP-based HIDS. As our focus is on the use of NLP, we consider HIDS that use syscall sequence and text data sources (e.g., text-based cyber threat intelligence data) as primary data source.

The key contributions of this review are as follows:
\begin{itemize}[]
    \item We are the first, to the best of our knowledge, to conduct a comprehensive and systematic review of the literature on the end-to-end pipeline of the use of NLP in HIDS development. 
    \item We propose a taxonomy for consistently classifying the current and future research on NLP-based HIDS. We demonstrate the use of our taxonomy by identifying 6 categories of NLP methods used in HIDS. For the evaluation of NLP-based HIDS, we categorize 133 attacks, 20 datasets and 17 evaluation metrics from the reviewed papers.
    \item We synthesize and discuss the pros and cons of each category of the used NLP methods, datasets, evaluation metrics and other key factors (e.g., classifiers, learning types) to help HIDS researchers and developers understand their characteristics.
    \item We highlight open issues with the current HIDS practices and propose potential research directions in the intersection of HIDS and NLP research.
\end{itemize}

Our SLR findings are expected to help in better understanding the landscape of the existing NLP-based HIDS for both HIDS researchers and HIDS developers. We recommend researchers to focus on developing real-time accurate HIDS leveraging advanced NLP techniques (e.g., text augmentation, NLP-based low-shot learning), enhancing HIDS model interpretability and semantic integration of HIDS in a Security Operation Center (SOC). For developers, our findings are expected to help them to enhance the existing HIDS or develop new HIDS using NLP method for supporting the construction and deployment of effective and efficient HIDS. To support developers in this regard, we highlight the relevant existing practices, considerations, advantages and limitations. We recommend that HIDS developers train and validate the HIDS model using NLP methods with their industry specific data before the deployment of HIDS as the most used public datasets that we found are usually outdated and lack sufficient and diverse attack instances.

\begin{table*}[t]
\centering
\caption{Comparative analysis of our SLR with the existing HIDS survey papers}
\label{tab:comp_survey}
\begin{threeparttable}
\resizebox{\textwidth}{!}{%
\footnotesize
\begin{tabular}{lcccccccccc}
\hline
 & NLP & Feature & Classifier & Attack &   & Evaluation  &  &  Common\\ 
Study &  Focus &  Extraction Tech &(ML/DL/Rules) & Types&  Datasets &  Metrics &  SLR & Papers\\
\midrule
Ours &  $\checkmark$  &  $\checkmark$ & All & $\checkmark$ (12)& $\checkmark$  &$\checkmark$ (17)  & $\checkmark$ & -\\ \hline
\citep{review_ids_4} &  $\times$  &  $\times$ & no DL & $\times$ & $\times$ & $\times$ &$\times$ & 2 \\ \hline
\citep{review_ids_5} & $\times$  & $\times$ &  no DL  & $\times$ & $\checkmark$  & $\checkmark$ (8) & $\times$ & 6   \\ \hline
\citep{review_ids_2} & $\times$  & $\times$ &  no DL  &  $\checkmark$ (4) & $\checkmark$ & $\times$ & $\times$ & 5  \\ \hline
\citep{review_ids_20} & $\times$ &  $\times$ &  $\times$ & $\times$ & $\times$ & $\times$  & $\times$ & 0  \\ \hline 
\end{tabular}}
\begin{tablenotes}
      \item\label{tnote:robots-r1} \footnotesize{\checkmark  represents yes, $\times$  represents no, -  represents not available}
      \end{tablenotes}
  \end{threeparttable}
\end{table*}

The rest of this paper is organized as follows. Section \ref{sec:preli} compares our SLR to existing surveys. Section \ref{sec:methodology} outlines our SLR methodology. Section \ref{sec:result} presents the overview of our results in terms of study distribution and our proposed taxonomy. Sections \ref{sec:RQ1} and \ref{sec:RQ2} present a detailed analysis of the used NLP methods for HIDS (RQ1) and evaluation of NLP-based HIDS (RQ2). In Section \ref{subsec:open_issues}, open challenges with future research directions are discussed. Section \ref{sec:threats} presents the threats to validity of our findings. Lastly, the paper is concluded in Section \ref{sec: concl}.

\section{Comparison to the Existing Literature Reviews}\label{sec:preli}
Intrusion Detection Systems (IDS) has consolidated its position as a must use defense technique with various commercial IDS to detect attacks within state, personal and industrial Information Technology (IT) infrastructures \citep{iot_ids}. IDS is a highly active research domain and the existing literature has been reviewed from diverse aspects due to the significance of intrusion detection to protect an organization from cyber attacks. For ensuring the unique and novel contribution of our SLR, we extensively analyzed the related reviews and compared them with our SLR in this section.

A large number of review studies \citep{review_ids_1, review_ids_11, review_ids_12, review_ids_13, review_ids_14, review_ids_3, review_ids_15, review_ids_21, review_ids_16, review_ids_7, review_ids_8, review_ids_9, review_ids_10, review_ids_17, review_ids_18, review_ids_19, review_ids_20} are available in the literature, which primarily focus on NIDS. NIDS is significantly different from HIDS as NIDS detects attacks in an organization's network, while HIDS detects attacks on an organization's host system after NIDS has failed or been bypassed \citep{review_ids_20}. While NIDS monitors the network traffic, HIDS monitors host data (e.g., syscalls). Compared to NIDS, HIDS has the ability to detect insider attacks and Advanced Persistent Threats (i.e, APT involves intruders staying in an organization's system for years by evading detection mechanisms) \citep{review_ids_5}. Due to the significant differences between HIDS and NIDS, the existing reviews \citep{svm_ids_survey, review_ids_1, review_ids_21, review_ids_3, review_ids_7, review_ids_8, review_ids_9, review_ids_10} have only 1 to 2 papers and the reviews \citep{review_ids_11, review_ids_12, review_ids_13, review_ids_14, review_ids_15, review_ids_16, review_ids_17, review_ids_18, review_ids_19, review_ids_20} have 0 papers in common with our HIDS focused SLR. 

Four reviews are available which focused on HIDS \citep{review_ids_2, review_ids_4, review_ids_5, review_ids_20} (comparative analysis shown in Table \ref{tab:comp_survey}). These reviews differ from our SLR in terms of objectives, included papers and results. None of these existing HIDS reviews focus on NLP. The semantics and contextual analysis ability of NLP methods help HIDS to detect unknown and adversarial attacks with lower False Alarm Rate (FAR) and higher accuracy \citep{s5, s30}. Hence, we focus on NLP to identify NLP-based methods (e.g., word embedding and language modeling) and the future research directions.

A decade-old review \citep{review_ids_20} focused on standalone HIDS and did not consider hybrid or collaborative HIDS, while our SLR does not confine the scope to standalone HIDS. The authors discussed network traffic, process behavior, file integrity and security of HIDS against tampering without focusing on any Machine Learning (ML), Deep Learning (DL) or NLP approaches. 
Another review \citep{review_ids_4} aims at the existing anomaly detection techniques of other domains and how they can be adopted from other domains to the HIDS domain, whereas our SLR discusses both misuse and anomaly detection-based HIDS. This review does not focus on features, attacks, datasets and evaluation metrics, which are discussed in our SLR. 
Another review \citep{review_ids_2} categorized the host data sources (e.g., system logs, windows registry) to discuss the existing literature from data source perspective. This review does not focus on the application of NLP methods, feature extraction methods (e.g., manual, automated), DL models and metrics, whereas our SLR analyses all these HIDS aspects. This review included the majority of the papers before 2010; also they prioritized the papers covering the use of host data sources including NIDS. In contrast, we conduct SLR in the (2010-2022) time-range as the adoption of NLP techniques in HIDS research gained momentum in this decade \citep{s30}. 
The review reported by \citep{review_ids_5} focused on syscall-based HIDS and their application on embedded systems, in contrast, we exclude (discussed in Section \ref{subsec:quality_assess}) the HIDS of any specific application area (e.g., embedded systems and IoT). 

\begin{table*}[]
\centering
\caption{Research questions that are addressed in our SLR}
\label{tab:RQs}
\footnotesize
\begin{tabularx}{\textwidth}{XX}
\toprule
Research Questions & Motivation\\
\midrule
\textbf{RQ1:} How have the NLP methods been used over the years by researchers to develop HIDS?
 & To identify and categorize the NLP methods and how these NLP methods are used to develop HIDS. The aim is to provide insights to HIDS developers and researchers about the existing prevalent practices and considerations in NLP-based HIDS and how NLP methods are beneficial to improve HIDS performance for industrial adoption.\\\hline
\textbf{RQ2:} How do researchers evaluate the NLP-based HIDS?\\
    $\bullet$ \textbf{RQ2.1:} What type of attacks do researchers aim to detect using NLP methods in HIDS? & To identify what type of attacks can be detected by NLP-based HIDS and what are the impacts of these attacks\\
    $\bullet$ \textbf{RQ2.2:} What are the datasets used to apply NLP methods in HIDS research? & To provide insight to practitioners and researchers on the HIDS datasets that are used to evaluate NLP-based HIDS including dataset type and availability.
\\
$\bullet$ \textbf{RQ2.3:} Which evaluation metrics are used to evaluate NLP-based HIDS? & To identify the evaluation metrics that are used to evaluate NLP-based HIDS\\
\hline
\end{tabularx}

\end{table*}

\textbf{In summary}, none of the above-mentioned existing HIDS reviews focus on NLP application, feature extraction methods and DL-based approaches, as shown in Table \ref{tab:comp_survey}. We identified, categorized and analyzed NLP methods, attacks, datasets and evaluation metrics used in NLP-based HIDS. Notably, none of the existing surveys covered DL-based techniques (e.g., language modeling, word embedding), while we identified 24 studies in the recent years using DL approaches with significantly better performance for real-life applications handling a huge volume of data. Considering the included papers, our SLR notably differs from these existing HIDS reviews having only 2, 6, 5 and 0 papers in common in the studies \citep{review_ids_4, review_ids_5, review_ids_2, review_ids_20}, respectively.

There are several reviews on NLP or text mining applications in different domains such as financial domain \citep{finance_nlp}, social network \citep{social_nlp}, market prediction \citep{market_pridct_nlp}, and a recent one in cyber security domain \citep{nlp_security}. The review in cyber security domain \citep{nlp_security} presents a broader view by identifying the tasks (e.g., cyber bullying, forensic analysis and sentiment analysis) in the cyber security domain which are supported by text mining, while our study thoroughly analyzes and focuses on NLP-based HIDS.

\section{Research Methodology}\label{sec:methodology}

We conducted a Systematic Literature Review (SLR) using the widely adopted SLR guideline \citep{slr_guidelines}. SLR is a widely adopted research approach in Evidence-Based Software Engineering (EBSE) \citep{kitchenham2004evidence} as SLR evaluates and interprets a research topic utilizing a reliable, rigorous and auditable methodology \citep{slr_guidelines}. For guiding our analysis, we aimed to answer two research questions (RQs) as shown in Table \ref{tab:RQs} with the corresponding motivations. Our review protocol steps are presented in the following Subsections (\ref{subsec:search_strategy}- \ref{subsec:data_extract_synthesis}).



\subsection{Search Strategy}\label{subsec:search_strategy}
We formulated our search strategy to retrieve the maximum number of relevant studies based on the guideline provided by \citep{slr_guidelines}. The search strategy includes the following steps.

\subsubsection{Search Method}
We used the automated database search method \citep{slr_guidelines} to retrieve the relevant studies from digital search engines and databases. We used the largest academic literature database Scopus digital library, which indexes over 5,000 publishers worldwide including the relevant sources (e.g., Elsevier, Springer) \citep{roland}. We complemented Scopus with IEEE Xplore and ACM Digital Library, which are the most frequently used academic digital libraries \citep{digital_lib}. Moreover, we complemented the automatic search by extracting more relevant studies using snowballing \citep{snowball_guidelines}.

\subsubsection{Search String}
We used the guidelines of \citep{slr_guidelines} to develop a comprehensive search string. Considering the key terms ``host intrusion detection" and ``NLP", we created several pilot search strings composed of synonyms and related terms. For the first term, we considered its varied representation (e.g., Host based intrusion detection, host IDS) along with the terms related to both anomaly and misuse detection. We excluded the term `HIDS' as it is also used to represent `high-dimensional and sparse (HiDS)', which provided irrelevant papers. Regarding the second term, we noticed the inclusion of `NLP' is not useful as the papers on HIDS may not specify such a term even though they use a wide variety of NLP techniques (e.g. n-grams, word embedding and language modeling). After executing a series of pilot searches in titles, abstracts and keywords of papers on databases and checking the inclusion of papers that were known to us, we designed the following search string.\\\\
\noindent\fbox{%
    \parbox{\linewidth}{%
        ``host intrusion detection" \textbf{OR} ``host based intrusion detection" \textbf{OR} ``host anomaly detection" \textbf{OR} ``host based anomaly detection" \textbf{OR} ``host based anomaly intrusion detection" \textbf{OR} ``host based ids" \textbf{OR} ``host ids" \textbf{OR}  ``host based misuse intrusion detection" \textbf{OR} (``signature based intrusion detection" \textbf{AND} ``host")
    }%
}
\begin{table*}[]
\centering
\caption{Inclusion and Exclusion Criteria}
\label{tab:IE}
\footnotesize
\begin{tabularx}{\textwidth}{X}
\hline
Inclusion Criteria\\
\hline
\textit{\textbf{I1:}} Publications that utilize NLP methods to develop host-based intrusion detection system including anomaly and misuse detection approaches.\\
\textit{\textbf{I2:}} Publications that are presented at a conference or a journal.\\
\textit{\textbf{I3:}} Studies that are published within the search timeline of January 2010 and May 2022 with extended coverage by snowballing until July 2022.\\
\hline
Exclusion Criteria\\
\hline
\textit{\textbf{E1:}} Papers that investigate IDS in a particular security application infrastructure (e.g., Cloud \citep{cloud_survey, cloud2}, IoT \citep{iot_ids} and Smart city \citep{smart_city_ids}).\\
\textit{\textbf{E2:}} Publications that are not written in English and are not accessible. \\
\textit{\textbf{E3:}} Unpublished papers that are uploaded to the archive, workshop papers and the conference version of a journal paper.\\
\hline
\end{tabularx}%
\end{table*}

\begin{table}[]
\centering
\caption{Quality assessment of the papers}
\label{tab:aq}
\footnotesize
\begin{tabularx}{\columnwidth}{lX}
\toprule
ID & Quality Assessment Criteria  \\
\midrule
AQ1 & Does the paper clearly state objectives or challenges it targets to solve in HIDS literature?\\
AQ2 & Is the proposed method well-defined and discussed in detail? \\ 
AQ3 & Is the performance of the proposed method measured and reported within the paper?  \\
AQ4 & Is the proposed method compared with other existing approaches? \\
AQ5 & Does the paper provide information regarding the datasets used for the assessment of the proposed approach? \\
AQ6 & Does the paper provide details of the varied evaluation metrics used for evaluation?   \\
\hline
\end{tabularx}
\end{table}

\subsection{Inclusion and Exclusion Criteria and Quality Assessment Criteria} \label{subsec:quality_assess}
Table \ref{tab:IE} shows the inclusion and exclusion criteria in line with our SLR aim and RQs. These are used for selecting the relevant papers out of the ones retrieved from data sources. 
We developed a quality assessment criteria for excluding low quality papers from our review. We adopted and adjusted our quality assessment criteria from \citep{aq2, aq3}. Table \ref{tab:aq} presents the Assessment Questions (AQs). We graded the reviewed papers on each quality assessment criterion using a three tier (``Yes"=1, ``Partially"=0.5 or ``No"=0) scale. The assessment score of a paper is calculated by adding the scores of the answers to the six AQs. To assure the reliability of our review's findings, we only included papers of acceptable quality, that is, those with an assessment score of more than 3.00 (50\% of the perfect score).



\begin{figure*}[]
  \centering
  \includegraphics[width=\textwidth]{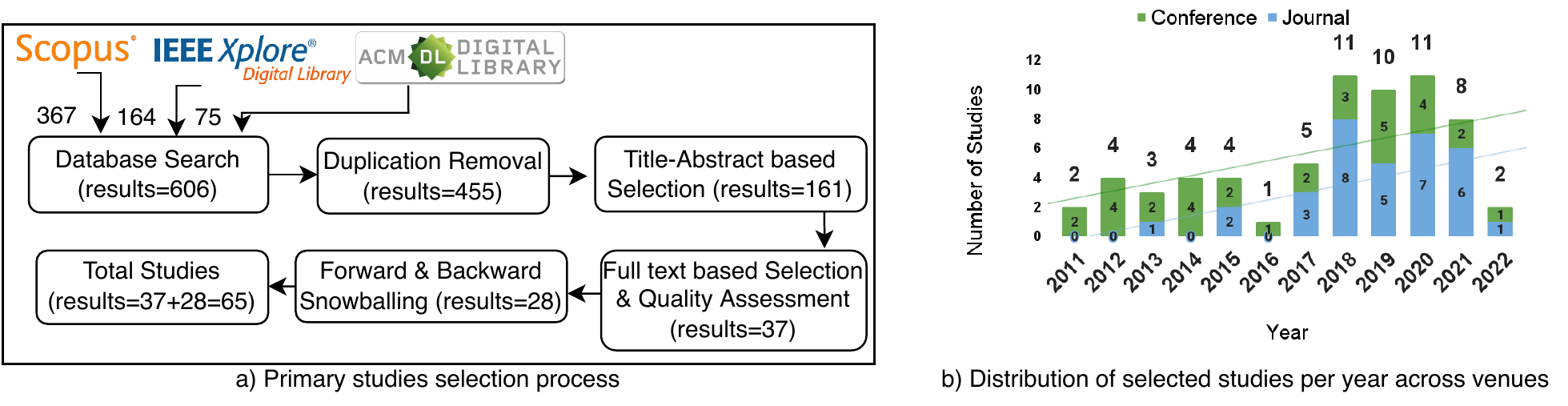}
  \caption{Primary studies selection process and distribution of selected studies.}
  \label{fig:paper_sel_dist}
\end{figure*} 
\subsection{Selection of Primary Studies}\label{subsec:study_selection}
Figure \ref{fig:paper_sel_dist} (a) shows the number of studies retrieved in different phases of the papers selection process. We performed \textbf{\textit{Database Search}}, \textbf{\textit{Duplication Removal}}, \textbf{\textit{Title-Abstract-based Selection}}, \textbf{\textit{Full-text-based Selection and Quality Assessment}} against the inclusion and exclusion criteria (Table \ref{tab:IE}) and quality assessment criteria (Table \ref{tab:aq}). Further, we used forward and backward \textbf{\textit{snowballing}} following the widely adopted snowballing guideline \citep{snowball_guidelines} by scanning the citations and references of the selected papers, respectively. Snowballing is used as the search string may not retrieve obscurely phrased studies, and the selected digital libraries may not exhaustively include all peer-reviewed papers \citep{snowball_guidelines}. For snowballing, we followed the same selection process including title and abstract, full text-based selection and quality assessment. In total 65 papers were selected for our SLR as enlisted in Appendix \ref{subsec:paper_list}, each with a unique identifier (S\#).

\subsection{Data Extraction and Synthesis}\label{subsec:data_extract_synthesis}
We organized a Data Extraction Form (DEF) based on 17 data items to answer our RQs as given in our online appendix \citep{online_appendix}. Data items (D1-D7) include context data such as title, author, venue, publication year, publisher, summary and open challenges, respectively. To facilitate the analysis of the extracted RQ relevant data, we grouped the data items as follows. RQ1 (D8-D12: NLP method, feature extraction method, learning type, classifier and HIDS type), RQ2.1 (D13-D14: attack detection/classification (i.e., binary or multi-class classification), attack instances), RQ2.2 (D15-D16: dataset, dataset availability) and RQ2.3 (D17: evaluation metric). A pilot study of 15 papers was conducted collaboratively by the authors to update the DEF for capturing the needed information in the best possible summarised form. The designed DEF was completed collaboratively using an online Google Spreadsheet and any ambiguity was resolved through discussion.


\textbf{\textit{Data Synthesis:}} We analyzed the context data items (i.e., D1-D5) using descriptive statistics as shown in Section \ref{subsec:studies_dist}. We analyzed the RQ relevant data items using thematic analysis considering the guideline of study \citep{thematic_ana} including the following steps. 
\textbf{\textit{Familiarizing with data}} by reading and examining our extracted data.
\textbf{\textit{Generating initial codes}} to capture NLP methods, attacks, datasets and evaluation metrics for NLP-based HIDS. 
\textbf{\textit{Searching for themes and generating potential themes}} for each data item by merging the corresponding initial codes based on their similarities.
\textbf{\textit{Reviewing themes and mapping themes}} were performed iteratively to review all the codes and themes to revise their allocations if needed. To finalize the RQs' answers, the synthesized results for each RQ were reviewed and any disagreements were discussed by the authors in daily Slack channel discussions and weekly meetings.
\begin{figure*}[]
  \centering
  \includegraphics[width=\textwidth]{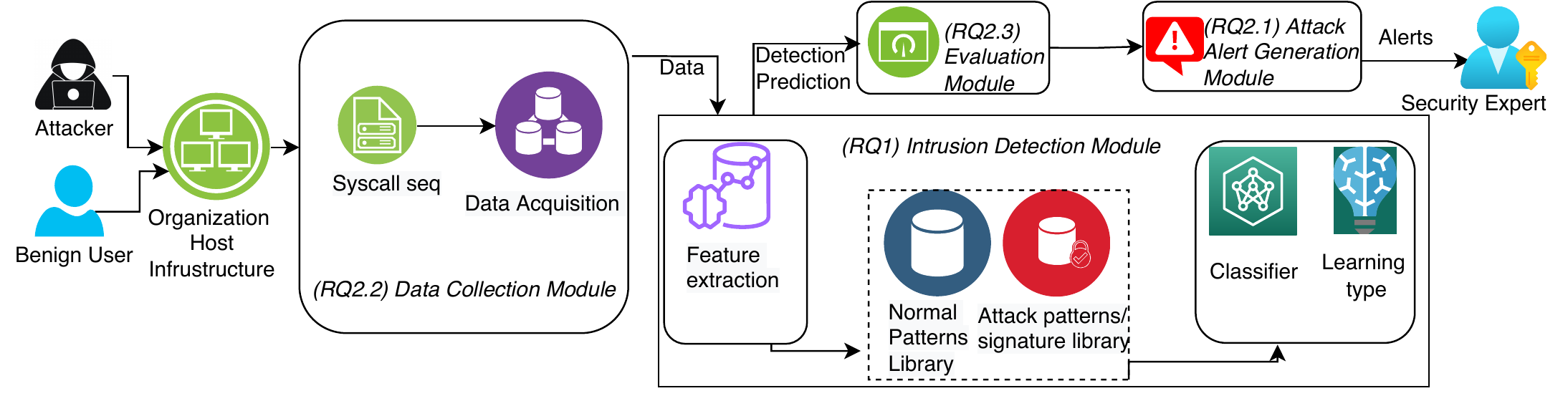}
  \caption{The main modules and a general overview of end-to-end pipeline of NLP-based HIDS}
  \label{fig:HIDS_overview}
\end{figure*}

\begin{figure*}[h]
  \centering
  \includegraphics[width=\textwidth]{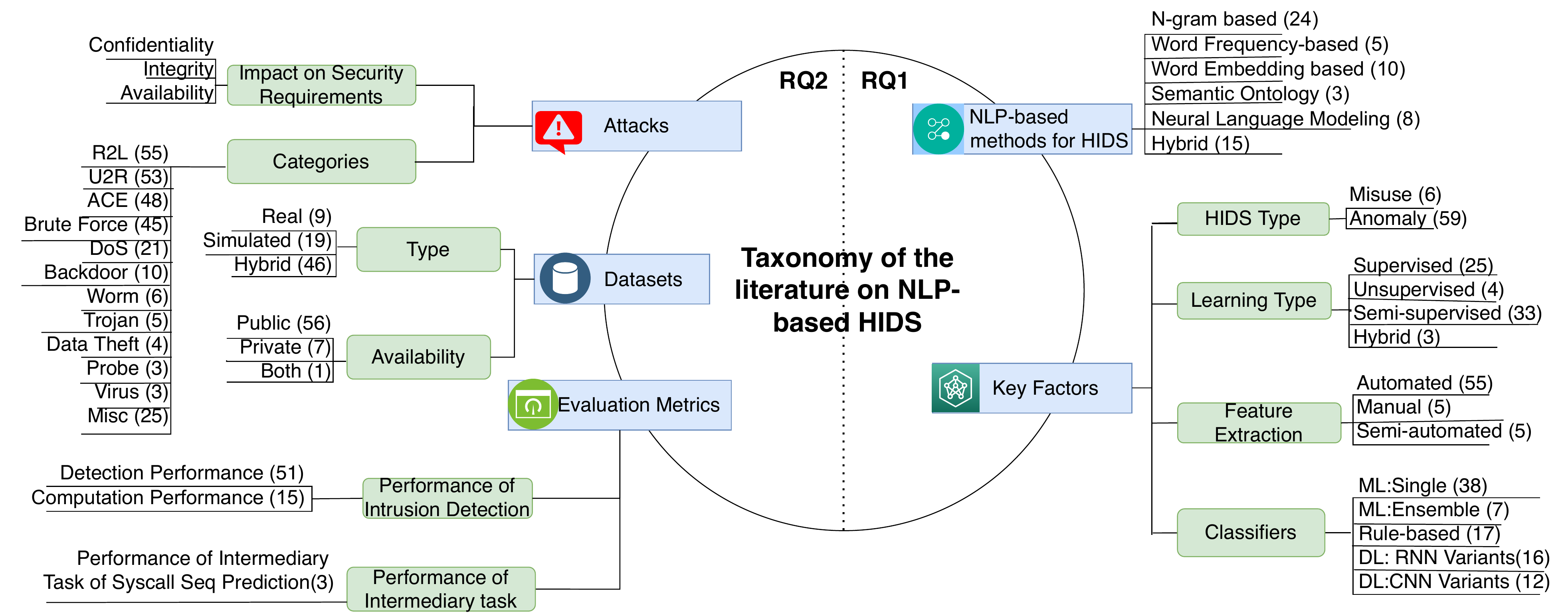}
  \caption{Our taxonomy of the literature on NLP-based HIDS with the relevant study counts}
  \label{fig:tax}
\end{figure*}

\section{Results Overview}\label{sec:result}
We present the distribution of the selected studies in Section \ref{subsec:studies_dist} and our taxonomy of the literature on NLP-based HIDS in Section \ref{subsec:overview}.

\subsection{Studies Distribution}\label{subsec:studies_dist}
Demographic information of papers (e.g., types) is considered helpful for novice researchers \citep{slr_mojtaba_cont}. Figure \ref{fig:paper_sel_dist} (b) presents the consistently upward trend of papers number in NLP-based HIDS in this decade due to the ever growing threat landscape. More than half of the total papers (42/65) and 81.82\% of the journal papers (27/33) were published between 2018 to 2022 indicating the rapidly growing research literature and maturity in adopting NLP in HIDS. 

The reviewed papers were primarily published in the following research areas: Cyber Security, Network Communications, Data Science and AI and also Software Engineering. The diversity of research areas in terms of the publication shows the interest of researchers with different research backgrounds in HIDS research.





\subsection{Our Taxonomy of the Literature on NLP-based HIDS }\label{subsec:overview}
Figure \ref{fig:HIDS_overview} shows the main modules and a general overview of an end-to-end pipeline of NLP-based HIDS. A HIDS typically monitors the events occurring in an organization's host infrastructure. First, the data collection module monitors and collects data (e.g., syscall traces), which capture valuable information about the running applications in a host system. 
Researchers consider syscall traces as documents, syscall sequences as sentences in the document and syscalls as words, which enables researchers to adopt NLP methods for HIDS in the intrusion detection module. In the intrusion detection module, the input syscall sequence data are processed and discriminative features are extracted using NLP methods from the data, which represent the normal or attack behavior of the input data. The automated NLP-based feature extraction methods help to generate reliable features, which accurately represent the behavior of a host system's activities. The generated features are usually applied to diverse classifiers (e.g., DL-based prediction models). These classifiers perform the attack detection. The evaluation module evaluates the attack detection performance of the intrusion detection module. Lastly, the attack alert generation module generates attack alerts based on the attack detection outcomes and shares them with the security experts of SOC in an organization.

Our SLR aims to cover the end-to-end pipeline of NLP-based HIDS and thoroughly discusses each of its modules. We propose a taxonomy of NLP-based HIDS by analyzing and synthesizing the extracted data from the reviewed papers. We provide a graphical representation of the NLP-based HIDS taxonomy in Figure \ref{fig:tax}. 

\begin{table*}[]
\centering
\caption{Number of papers in NLP method categories used in HIDS which are published in each year from 2011 to 2022 }
\label{tab:det_year}
\footnotesize
\begin{tabularx}{0.75\textwidth}{lrrrrrrrrrrrrr}
\toprule
Category	& 11	& 12	& 13	& 14	& 15	& 16	& 17	& 18	& 19	& 20	& 21	& 22	& \textbf{Total}\\
\midrule
Word frequency-based 	& -	& -	& 2	& 2	& -	& -	& -	& -	& -	& -	& 1	& -	& 5\\
N-gram-based 	& 1	& 3	& 1	& 2	& 4	& 1	& 4	& 3	& 2	& 3	& -	& -	& 24\\
Word embedding-based	& -	& -	& -	& -	& -	& -	& -	& 1	& 1	& 3	& 3	& -	& 8\\
Neural language modeling & -	& -	& -	& -	& -	& -	& -	& 2	& 3	& 1	& 2	& -	& 8\\
Semantic ontology & -	& 1	& -	& -	& -	& -	& -	& 2	& -	& -	& -	& -	& 3\\
Hybrid	& 1	& -	& -	& -	& -	& -	& 1	& 3	& 4	& 4	& 2	& 2	& 17\\
\textbf{Total}	& 2	& 4	& 3	& 4	& 4	& 1	& 5	& 11	& 10	& 11	& 8	& 2	& 65\\
\hline
\end{tabularx}
\end{table*}

In RQ1, we identify and categorize the NLP methods (e.g., word embedding, semantic ontology) used in HIDS. We also focus on and categorize each key factors that are considered in the intrusion detection module. The key factors that are adopted to investigate the host data for intrusions include HIDS type (e.g., misuse, anomaly), learning types (e.g., supervised, unsupervised), feature extraction techniques (e.g., manual, automated) and classifiers (e.g., ML, DL). RQ1 including the categorizations of NLP methods and all the key factors is discussed in detail in Section \ref{sec:RQ1}.

In RQ2, we analyze how the NLP-based HIDS are evaluated in terms of detected attacks (in RQ2.1); used datasets (in RQ2.2) and used evaluation metrics (in RQ2.3). In RQ2.1, we categorize the attacks (e.g., Brute Force, Backdoor) that can be detected by leveraging NLP-based HIDS for which alerts are generated in the attack alert generation module. We also focus on the impacts (e.g., confidentiality, integrity) of these attacks on the security requirements. The attacks and impacts are discussed in detail in Section \ref{subsec:attacks}. In RQ2.2, we identify and analyze the datasets that are used in the data collection module for applying NLP methods to develop and evaluate HIDS. We also focus on the data types (i.e., real, simulated) and their availability (i.e., private, public). Section \ref{subsec:dataset} details the datasets that are used in the reviewed studies to evaluate NLP-based HIDS. Lastly, in RQ2.3, we identify and categorize the metrics (e.g., FAR, Detection Rates (DR)) used for the evaluation of HIDS in the reviewed literature. Detailed comparative analysis of the evaluation metrics is presented in Section \ref{subsec:metrics}. 

We reflect on the prevalent practices based on findings to our RQs as shown in the taxonomy of Figure \ref{fig:tax}. To help researchers and developers to identify and select their choices, a comprehensive comparative analysis in terms of strengths and weaknesses for the categories of this proposed taxonomy are highlighted in Tables 7, 8, 10, 11 and 12. For conciseness, we added the study mapping with the corresponding categorization of the RQs in our online appendix \citep{online_appendix}. For the frequently used acronyms in this article, Table \ref{tab:abbr} in Appendix \ref{subsec:notation} presents their abbreviations.

\section{RQ1: Use of NLP Methods in HIDS} \label{sec:RQ1}
Firstly, we categorize and discuss different NLP methods that researchers use in HIDS, and how the use of these NLP methods in HIDS trend is evolving over the years in Section \ref{subsec:nlp_cat}. To highlight the effectiveness of the NLP methods for HIDS, we only cover the NLP methods proposed by the reviewed studies and disregard the baseline approaches (i.e., methods used to compare the effectiveness of the proposed method). For the development of HIDS using different NLP methods, HIDS developers need to consider a number of other factors (e.g., learning type and classifier). Hence, these key factors that are considered for developing HIDS using NLP are summarized in Section \ref{subsec:key_fact}.


\subsection{Categorization of NLP Methods Used in HIDS}\label{subsec:nlp_cat}
We present a comprehensive and novel categorization of NLP methods used in HIDS. Table \ref{tab:det_year} demonstrates the distribution of the identified categories of NLP methods used in HIDS over the years. Table \ref{tab:det_year} suggests that the prevalent NLP methods used in HIDS over the last 4 years are word embedding, language modeling and hybrid (i.e., combine multiple NLP methods). The increasing amount of large scale security data and evolving complex attack patterns have contributed to this trend of using these NLP methods that are evolved over deep learning models due to their prominent HIDS performance. Table \ref{tab:tax} shows the categories along with their advantages, disadvantages and mapped studies. The categories of used NLP methods in HIDS are discussed below.

\begin{table*}[]
\centering
\caption{Categories of NLP methods in HIDS along with their strengths, weaknesses and mapped studies}
\label{tab:tax}
\footnotesize
\begin{tabularx}{\textwidth}{p{0.09\textwidth} p{0.45\textwidth} p{0.23\textwidth} p{0.15\textwidth}}
\toprule
\textbf{NLP Method} & Strengths & Weaknesses & Study Ref\\
\midrule

Word freq-based & 
$\bullet$ Less computational and memory overhead\newline  
$\bullet$ Lightweight and easy to dynamically update normal host system behavior profile 
& $\bullet$ Do not consider the order of syscalls \newline 
$\bullet$ Difficult to gain a good detection rate and FAR compared to n-gram & S28, S29, S31, S32, S55 \\\hline

N-gram-based & $\bullet$ Holds order of syscalls to some extent\newline 
$\bullet$ Enables early attack detection by splitting the process trace of a host system into smaller sequences \newline 
$\bullet$ Achieves good detection and FAR than word freq-based method \newline 
$\bullet$ Reduces the possibility of mimicry attacks on HIDS {[}S59{]} 
& $\bullet$ Incur high computational and memory overheads \newline 
$\bullet$ Requires longer training time compared to Word freq-based method & S2, S11, S16, S17, S19, S23, S24, S25, S27, S30, S34, S38, S39, S41, S42, S49, S52, S53, S56, S58, S59, S60, S61, S65 \\\hline

Word embedding-based  & 
$\bullet$ Word embedding learns hidden semantic relations of syscalls. \newline 
$\bullet$ Word embedding gives dense vector representation compared to One-hot encoding that requires sparse and high dimensional vectors. \newline 
$\bullet$ Word embedding has better generalization ability than n-gram-based methods {[}S57{]}. \newline 
$\bullet$ Can help in domain adaptation through transfer learning to mitigate the requirement of huge domain-specific data {[}S12{]} 
& $\bullet$ Word embedding requires complex DL-based training \newline 
$\bullet$ Requires more computation cost compared to n-gram or TF-based methods & S4, S12, S14, S20, S46, S47, S51, S57 \\\hline

Semantic ontology & 
$\bullet$ Enables attack knowledge fusion from heterogenous textual data sources (e.g., security bulletin) through a common semantic schema \newline 
$\bullet$ Link and infer means and consequences of cyber threats and vulnerabilities whose signatures are not yet available \newline 
$\bullet$ Provide HIDS semantic expressiveness and knowledge description 
& $\bullet$ Expert knowledge is required to identify the data sources and define the detection rules. \newline 
$\bullet$ Manual creation and update of detection rules are required. & S13, S26, S33 \\\hline

Neural lang model & 
$\bullet$ Predicts a future syscall sequence possibly to be executed during an attack. \newline 
$\bullet$ Combining the known invoked syscall traces with predicted future syscall sequences helps to improve the intrusion detection performance \newline 
$\bullet$ Modeling syscalls helps to capture interword relationships. 
& $\bullet$ Modeling the system behavior requires a huge amount of data  & S5, S15, S21, S22, S43, S44, S50, S63, \\\hline

Hybrid & 
$\bullet$ Makes HIDS more reliable and resilient against evasion and adversarial attacks by combining decisions from heterogeneous detectors \newline 
$\bullet$ Gains the advantages of multiple features or models to lower FAR 
& $\bullet$ Requires high computation overhead compared to single methods. & S1, S3,S6, S7, S8, S9, S10, S18, S35, S36, S37, S40, S45, S48, S54, S62, S64\\\hline
\end{tabularx}
\end{table*}

\subsubsection{\textbf{Word frequency-based}}
In NLP, the word frequency-based method considers how frequently a word (i.e., one syscall) occurs within a document to indicate the significance of a specific word within the overall document, which is a lightweight method. As the normal behavior of the host varies over time, dynamic update of normal behavior is critically challenging [S31]. Hence, a lightweight method like the word frequency-based method is required to dynamically update normal behavior and extract the features in HIDS with less computation cost. Researchers investigated the use of word frequencies considering \textbf{number of syscalls (n)=1}. Word frequencies are adopted by using three commonly used approaches in NLP, which are Term Frequency, Inverse Document Frequency and Bag-of-Words. Though these word frequency-based methods are computationally lightweight, they are less detailed as they do not hold sequence information, which makes it challenging to gain good detection rate and FAR [S9, S31].

\textbf{Term Frequency (TF):} TF is the calculation of how frequently a term occurs within a document. As each syscall is presented by a unique word, consequently, the syscall sequences become sentences and syscall traces holding the syscall sequence denote documents. To enforce stronger separability between normal and abnormal syscalls, researchers used TF in One-Class Support Vector Machine (OC-SVM) [S29] and Clustering with Markov Network (CMN) [S32] classifiers. To reduce the dimension of TF vectors into a lower dimensional space, researchers successfully used principal component analysis (PCA) [S28] that reduced the computational cost.
 
\textbf{Inverse Document Frequency (IDF):} In NLP, IDF calculates if a word is common or rare in a given document corpus. IDF is calculated by the logarithm of the quotient that is obtained by the division of total document count and count of documents that contain the word. A study [S31] focused on the frequencies of the syscalls involved in a trace to generate three feature vectors (i.e., TF, IDF, standard deviation of IDS) and applied them to K-Nearest-Neighbours (KNN). Though the diverse syscall frequency-based classifier gained an acceptable performance for a few types of attack, it can not completely realize the complex behavior of a modern host system, consequently, require more intelligent HIDS.

\textbf{Bag-of-Words (BoW):} Another commonly used word frequency-based technique is BoW, which represents text by describing words' occurrence within a document. BoW discards words' order or structure related information in a document. BoW represents if words are present in a document, not the words' location in the document, and does not focus on the contextual relation of words in the document. To better identify multiple attack types, BoW is used for characterizing the syscall-based HIDS data to be used in Deep Belief Networks (DBN)-Softmax DL method for achieving high accuracy [S55].

\subsubsection{\textbf{N-gram-based}}
N-gram is one of the most used traditional NLP methods for representing sentences that holds the context information [S5]. While sentences are sequence of words, syscall sequence is a sequence of syscalls. Due to the similarity of the representation of sentences and syscall sequence, n-gram is highly used in HIDS, where n-gram refers to a sequence of n (\textbf{n$>$1}) syscalls extracted from a syscall trace [S41]. N-gram method extracts sequence of syscall of length `n' from a syscall trace by sliding a window one syscall at once. For instance, a syscall trace with syscalls ``open, getrlimit, mmap, close” [S39], two sequences ``open, getrlimit, mmap” and ``getrlimit, mmap, close” of length 3 (n=3) can be extracted. As n-gram holds the order of syscalls, n-gram has the ability to preserve sequence information of the syscall to provide good detection accuracy and it reduces the possibility of mimicry attacks on the HIDS [S59]. Besides, a HIDS should intervene early to stop the exploited process instead of waiting for the exploit to complete. To enable early attack detection, n-gram is used to split the syscall trace of a process into smaller sequences which can be analyzed instead of waiting for the process to complete. Due to these advantages, n-gram, frequency of n-grams, bag-of-n-grams and lookahead pairs methods from this n-gram category are used to develop HIDS.

\textbf{N-gram:} A set of studies [S16, S23, S24, S52] have used n-gram to extract sequence from benign syscall traces and compare them with sequences of a HIDS input trace. Then, any new sequence in the HIDS input trace is detected as anomalous. These studies adopted different similarity measures commonly used in NLP to measure similarity among the sequences such as Hamming distance [S23, S24], Sequence Covering for Intrusion Detection (SC4ID) [S16], Levenshtein’s distance (LEV) [S16], Longest Common Subsequence (LCSq) [S16], Longest Common Substring (LCSt) [S16] and improved edit distance [S52]. These n-gram-based methods, which use similarity measures are highly dependent on the similarity threshold for anomaly detection. Identifying the suitable threshold requires domain expertise and can be error-prone. To mitigate dependence on threshold values and to handle large scale syscall-based raw input data, in recent studies, n-gram is used in advanced DL models (e.g., Long Short Term Memory (LSTM) [S11]), Convolutional Neural Network (CNN) [S17] and Variational Auto Encoder (VAE) [S38]). For example, a study [S11] used n-gram (i.e., as syscall sequences are of variable length) to detect known and unknown attacks. This study [S11] modeled the normal behavior of the host system by syscall using LSTM architecture to detect zero-day \citep{zero_day_attack} unknown attacks. Then, to predict the most probable `n’ syscalls during the attack detected by the LSTM, the Multiple Hidden Markov Model (Multi-HMM) is adopted. Here each Hidden Markov Model (HMM) is used to model a distinct known attack for predicting the probable ‘n’ syscalls during the corresponding attack.

\textbf{N-gram frequency:} Although n-gram preserves the syscall traces' sequential information, the representation vectors are long. Hence, with an increasing value of n, the detection model requires more storage space and processing time [S2]. To mitigate this issue, a few studies [S19, S34, S65] used only the most frequent n-grams to reduce the feature space at the expense of the loss of some relevant information, which is a common NLP method. Though extracting syscall frequency is fast, syscall count in the trace is not sufficiently informative for detecting anomalies as syscalls' count can be similar in normal and abnormal syscall traces. To mitigate this issue, the n-gram frequency (i.e., phrase count) of NLP is used by a set of studies [S2, S25, S30, S41, S53, S59, S60] for obtaining a better detection rate with faster processing. In this method, distinct syscall sequences of varying lengths are counted. For example, to adopt n-gram frequency, a study [S41] extracted n-grams of syscalls by sliding a window of `n' syscalls over the syscall trace, associated a feature vector for each n-gram, and calculated each n-gram's occurrence frequency for diverse traces. These varied-length n-grams are represented as fixed-size vectors and their cumulative occurrence frequency is assigned as weight. As this method holds temporal order of syscall in a trace, the application of OC-SVM trained on n-gram frequency showed better detection accuracy than the HIDS methods using TF and TF-IDF.

\textbf{Bag-of-n-grams:} The ability to preserve the relative order of syscalls makes the bag of n-grams a better alternative to the Bag-of-Word (BoW) NLP method in the context of syscall-based HIDS. BoW only considers the occurrence frequency of a word in a document to determine how much the document is relevant to specific words. In the context of syscall data, considering syscall as a word and syscall trace as a document, the BoW model can not preserve the relative order of syscalls. For example, the feature vector using BoW of syscall traces “S1: open, getrlimit, mmap, close” and “S2: open, mmap, getrlimit, close” are similar. However, to model process behavior, the relative order of syscalls is highly significant. Loss of information of syscall sequence may leave a host vulnerable to the mimicry attack [S42], where an attacker interleaves malware syscall trace patterns with benign syscall traces. Thus, bag-of-n-grams consider multiple consecutive syscalls as one term, which helps hold the sequence information and prevent mimicry attacks. A study [S42] showed that bag-of-n-grams help more accurately distinguish process behavior through syscalls for intrusion detection in HIDS.

\textbf{Lookahead pairs:} Another variation of the n-gram NLP method is lookahead pairs. In the lookahead pairs method, for a given syscall sequence and a window of size (n), this window is slid over the syscall sequence. Then, except for the first syscall of the window, each syscall of the window is paired with the first syscall of that window. For example, for a syscall sequence ``S1: open, read, close, exit". For n=3, starting at the first syscall ``open’’ the window now has 3 elements ``open, read, close”. Next, the pairing process pairs as follows: ``open, read", ``open, close". Then, sliding the window next syscall (i.e., read) and the pairing process pairs as follows ``read, close" and ``read, exit". Hence, the lookahead pair method presents a model of constant space complexity for syscall-based HIDS. A study [S56] showed that lookahead pairs modeled with their occurrence frequency are good discriminators of benign and malicious program executions and is adversarial attack tolerant in terms of some contamination in the training data.

\subsubsection{\textbf{Word embedding-based}}
Word embedding is a sequential representation method to capture semantic relation and contextual information from syscall with reduced size vector. Word embedding has better generalization capability than n-gram-based methods [S57]. Word embedding significantly enhances the NLP classification task's performance due to capturing semantic features that motivates HIDS researchers to use word embedding to deal with syscall sequence [S57]. While One-Hot Encoding (OHE) in NLP is the simplest method to represent categorical/text data for a classifier [S9]. In HIDS, to use OHE, each syscall (word) is presented as a vector, where each position in the vector represents a specific syscall. Hence, input vector size is the number of diverse syscalls. A specific syscall is mapped to a binary vector with all zeros except for a single one at the suitable position for the corresponding syscall. OHE does not require specific prior knowledge and complex feature selection [S8]. Additionally, it preserves the original information of the data intact [S8]. However, using long OHE representation for syscalls can lead to training neural networks with a huge number of weights [S9]. In NLP, to mitigate the computational complexity and high training time due to sparse and high-dimensional vectors of OHE, word embedding has proven its value. Hence, it is preferred to use shorter vectors generated by word embedding to represent syscalls. To detect intrusion in HIDS, researchers use diverse word embedding methods from NLP such as Word2Vec [S14, S47, S51], GloVe [S47, S51], FastText [S51] and Graph Random State Embedding (GRSE) [S20] to represent the syscalls. 

\textbf{Word2Vec} \citep{word2vec_mikolov} in NLP generates word vectors based on the context in which they are used. Word2Vec depends on words' local context, in contrast, \textbf{GloVe} \citep{glove} uses global statistics on word co-occurrence for learning words' vector representations. In NLP, GloVe is a count-based model that learns a word vector representation using co-occurrence information, i.e., for each word, GloVe count how frequently those words are used in some context in a large corpus. \textbf{FastText} \citep{fastText_mikolov} developed by Facebook AI research extends Word2Vec by considering sub-word information, whereas Word2Vec and GloVe do not consider sub-word information. As the syscall names are chosen based on their functionality, sub-words can provide supplementary information about relationships among syscalls [S51]. For instance, directory is represented as `dir’ in short. Syscalls including this sub-word (e.g., cddir, mkdir, rmdir, readdir, fchdir, mkdirat) usually represent changes in the host's file system kernel module [S51], which is important for profiling host behavior. Moreover, such sub-words include supplementary information about their context (e.g., ccdir and mkdir represent that they handle a directory). 

Syscall sequences are essentially considered as word sequences (i.e., sentences). However, syscall sequence has several distinct features compared to sentence. For instance, syscalls have an implicit pattern relation, and the changing order among patterns has less effect [S20]. Further, syscall occurrence count often has little effect on the detection outcome. Syscalls also have other characteristics (e.g., resource utilization, execution time, predefined rules and empirical weights of syscalls). The NLP methods that are usually used in HIDS (e.g., BoW, TF-IDF, n-gram and Word2Vec), do not completely exploit such relations in syscall sequence and can not simply support such properties [S20]. To mitigate these limitations, a sequence embedding approach, namely, \textbf{Graph Random State Embedding (GRSE)} is proposed using Word2Vec [S20]. To generate vectorized representation of graph nodes GRSE performs random walk over the graph network for generating graph node paths, which emulates the text generation process (for HIDS, i.e., the process executing a syscall) for obtaining various graph node access sequences. Further, GRSE utilizes Word2Vec for learning all the nodes' vector representations based on the random walk sequence, which improves a HIDS's performance.

A major challenge to develop HIDS with better performance for a host domain with low resource is to mitigate the requirement of huge domain-specific data. To mitigate this challenge, word embedding helps in domain adaptation following NLP approach through transfer learning. A study [S12] showed the applicability of word embedding-based transfer learning by adopting a word embedding-based LSTM model that uses a significantly lower amount of target HIDS domain data to detect attacks in multiple host domains. This study [S12] further discussed both homogeneous and heterogeneous transfer learning-based HIDS problems, where monolingual word embedding is required for the homogeneous HIDS domain and bilingual word embedding is needed for the heterogeneous HIDS domain.

\subsubsection{\textbf{Semantic ontology}} To develop a HIDS with the capability of potentially linking and inferring means and consequences of cyber threats and vulnerabilities whose signatures are not yet available, knowledge fusion from heterogeneous textual data sources is required [S33]. For example, cyber threat intelligence data are available at various textual sources (e.g., blogs and security bulletins) [S26]. However, to gain the advantages of heterogeneous data sources to detect any threats, a common semantic schema is required to integrate information from disparate sources (e.g., different concepts and standards such as STIX, CVE, CVSS, CAPEC and CYBOX from diverse intelligence sources) [S33]. To mitigate this issue, a semantic ontology in NLP can provide a common semantic schema for integrating information from diverse sources by describing concepts and relationships among the concepts. To improve HIDS performance, semantic ontologies are used based on a semantic-reasoner along with a rule engine for detecting intrusion [S13]. Semantic ontology not only easily integrates information from textual sources but also helps to detect existing attacks' variations. For evading attacks, attackers often use diverse tools which can perform similar activities or use combinations of tools/techniques used in older attacks. Further, attackers often use similar tools and exploit similar vulnerabilities in varied attacks. As semantic ontology can integrate information of diverse attacks and fuses it with textual information, semantic ontology-based HIDS can detect such new attacks [S26].

The first ontology-based HIDS [S13] aimed at providing semantic expressiveness and knowledge description for improving IDS performance and reducing the search time for malware scanning. This study [S13] developed HIDS using Semantic Web technologies that give well-defined meaning for enabling computers and human to work in cooperation. Later, a study [S26] proposed continuous integration of information from dynamic textual sources (e.g., threat intelligence information of attack patterns, prior attacks, tools used for attacking and indicators) and combined it with malware behavioral information for detecting known and unknown attacks. This study [S26] used an off-the-shelf \textbf{Named-Entity Recognizer (NER)} from NLP trained on cyber security text to extract entities from plain text. Similarly, another study [S33] extracted named entities from the unstructured web-text for the continuous evolution of ontology. This study [S33] integrated text data from the web with different sensor streams, domain expert knowledge and constructed the ontology based on three main classes: ‘means’, ‘consequences’ and ‘targets’. The ‘means’ class captures the approaches used to execute an attack, the ‘consequences’ class captures the results of the attack and the ‘target’ class captures the details of the host under attack. Finally, this study [S33] has demonstrated the effectiveness of the semantic integration of web text for detecting threats in HIDS. 

\subsubsection{\textbf{Neural language modeling }}
To capture inter word relationships and predict the next words in NLP, a neural language model performs probability distribution calculation over words' sequences. Neural language modeling has achieved significant performance in realizing real-world NLP tasks [S15], such as Google’s autocompletion and voice assistant. Inspired by such significant performance, HIDS researchers discovered surprising resemblance between the host’s syscall sequence and natural language [S44, S63]. Researchers consider syscall sequences as instances of the language for communication between users (or programs) and host, where syscall and syscall sequences refer to words and sentences in natural languages [S63]. Due to this high resemblance of syscall sequence with natural language, researchers adopt neural language modeling in HIDS to model syscall. 

To monitor host system state and predict attack behavior, a language model-based prediction module predicts syscall sequence which will be executed in the future (i.e., possibly to execute during a known or unknown (zero-day) attack) [S44, S63]. The use of neural language modeling enables HIDS to model the semantic meaning of syscalls and analyze sequences at the sentence level for building a robust syscall Sequence-to-Sequence (Seq2Seq) prediction model. The Seq2Seq prediction model is adopted from NLP Question-Answer model for generating future syscalls considering prior invoked syscall sequence as question and generated syscall sequence as answer [S63]. Further, combining the prior invoked syscall traces with the predicted syscall sequence drastically enhances the intrusion detection performance, which is verified using different classifiers in HIDS [S15, S44]. As a Recurrent Neural Network (RNN) has the inherent capability to process sequential data and can memorize previous results, diverse variations of RNN (e.g., LSTM and Gated Recurrent Units (GRU)), RNN Variational Encoder-Decoder (VED) and combination of CNN with RNN-based language models (e.g., CNN-GRU, LSTM-textCNN) are adopted to handle sequential syscall data in the HIDS research domain.

\textbf{RNN-VED-based language model:} Seq2Seq language model based on variational encoder-decoder (VED) and variants of RNN showed promising performance in NLP. Researchers [S5, S43] exploited the semantics behind the invoking order of syscalls that are considered sentences. The RNN-VED in HIDS [S5, S43] learns the correlation between the syscalls to predict the future possible syscall sequence based on the given context. Another study [S63] used RNNs and Variational Autoencoding language modeling for learning long-term syscall sequences executed during an attack. This approach enables effective prediction and classification of syscall sequences that are possibly to occur next during known or unknown (zero-day) attacks.

\textbf{LSTM-based language model:} Evolutionary intrusion attacks (e.g., obfuscation technique) can change the malicious syscall sequence so that it can bypass the intrusion detection; and still can gain the same invasion purpose and effect that makes the detection outcome not robust and even invalid. To detect obfuscation attacks, a study [S22] used a behavioral semantics enhancement method by using an LSTM Seq2Seq language model that performs sequence completion. Then, the enhanced sequences are represented as vector matrices which are the input for the multi-channel Text-CNN that makes the model robust to obfuscation attacks. This language modeling-based method showed better intrusion detection performance and robustness than other NLP methods such as syscall combinations, n-gram subsequences and feature matching frequencies. LSTM model in NLP requires a long training time and has a complex structure. With the motivation of reducing training time, a study [S50] used NVIDIA CUDA Deep Neural Network (CuDNN) library to adopt CuDNNLSTM language model which is supported only on GPU systems. CuDNNLSTM language model using a bidirectional encoder and a unidirectional decoder is approximately ten times faster than LSTM model for faster convergence during training for HIDS development [S50].

\textbf{GRU-based language model:} The structure of GRU is simpler and GRU can reduce the required training parameters compared to LSTM. GRU [S44] and CNN-GRU [S15] language models are used in the literature for reducing the required training time and improving the efficiency of HIDS. In the CNN-GRU language model [S15], CNN layers are able to capture syscalls' local correlations in the sequence and can improve efficiency by executing in parallel. Then, the GRU layer is able to learn syscall's sequential correlations from these high-level features. The CNN-GRU language model gained near state-of-the-art performance and substantially reduced training time compared to LSTM models. Traditional HIDS models consider syscall traces generated by an individual process. However, multiple processes are often utilized by modern applications and one or more of these processes are impacted by modern attackers. Hence, to modernize HIDS a Wave-Net architecture is used to aggregate predictions for all processes corresponding to an application [S21]. This approach [S21] outperformed the CNN-GRU language model [S15].  

\subsubsection{\textbf{Hybrid }} To gain the advantage of multiple NLP methods, researchers have adopted hybrid methods by combining different NLP methods. The hybrid method includes the combination of heterogeneous classifiers using different NLP methods, a combination of n-gram and TF-IDF, a combination of n-gram and statistical approaches and a combination of n-gram and data augmentation methods.

\textbf{Combination of heterogeneous classifiers using different NLP methods:} To reduce the FAR of HIDS and make the HIDS more reliable and resilient against evasion and adversarial attacks, a set of studies effectively combined decisions from heterogeneous classifiers by using different NLP methods. A few studies [S35, S64] have trained Sequence Time-Delay Embedding (STIDE) and HMM utilizing n-gram-based sequential features, and OCSVM utilizing TF-IDF vectors and effectively combined decisions from these heterogeneous classifiers. Another study [S66] used a deep Multi-Layer Perceptron (MLP)-based Neural Network (NN) to combine the detection outputs of STIDE (n-gram-based), text classifier (TF-based) and syscall graph-based detection. The combination of heterogeneous classifiers with heterogeneous NLP methods showed consistently better performance than a single detector and ensemble of homogeneous detectors [S35]. 


\textbf{Combination of n-gram and TF-IDF:} TF-IDF denotes the multiplication of TF and IDF scores of the word that is used to weigh down the frequent words and scale up the rare ones. A set of studies [S3, S6, S7] used TF-IDF vectorized n-gram vectors to include sequence information to some extent (via n-gram) for improving detection performance with lightweight TF-IDF-based feature calculation for computation cost efficiency. Being inspired by sentiment analysis in NLP, where combining 2-gram, 3-gram and 4-gram models improved accuracy, a study [S3] combined 2-gram, 3-gram and 4-gram TF-IDF from selected syscalls of a trace file. Based on these combined features, for early detection of intrusions, researchers built a HIDS which processes the first few hundred syscalls to detect intrusion for an application. Since software applications' complexity has increased drastically, such applications invoke a large number of syscalls in a short time, which demands early detection that can consequently reduce HIDS resource consumption. Further, truncated Singular Value Decomposition (SVD) based on the TF-IDF values is used for reducing the transformed n-gram feature vectors' dimensionality, which gained high detection performance with low processing overhead [S6].

\textbf{Combination of n-gram and statistical approaches:} Diverse statistical approaches to the n-gram-based NLP method have been adopted in HIDS research for OS independent feature generation. Such OS independent features have several advantages such as (i) model transferring: HIDS model trained on data from OS X can be utilized on OS Y as the data share the same feature subspace and (ii) data transferring: data from OS X can be selectively combined with the data from OS Y to enrich the training data [S1]. Statistical methods used in HIDS are standard deviation [S1, S40], mean [S1], skewness [S1], kurtosis [S1], standard error [S1], eigenvector [S37] and entropy [S1, S36]. For example, a study [S1] showed that statistical methods (e.g., standard deviation and skewness) on n-gram frequency outperformed the following NLP methods: (i) bag-of-syscalls; (ii) TF-IDF (iii) n-gram frequency; and (iv) n-grams based methods.

\textbf{Combination of n-gram and data augmentation methods:} HIDS datasets usually suffer from data imbalance problem as gaining malicious syscall trace is challenging and much less in number compared to benign trace. The data imbalance problem creates a bias in the HIDS prediction model towards the majority benign class. To mitigate the data imbalance problem, a set of studies used different approaches on the n-gram NLP method such as Synthetic Minority Over-sampling Technique (SMOTE) [S40] and data augmentation using Generative Adversarial Nets (GAN) [S48], NLP-based Sequence Generative Adversarial Nets (SeqGAN) [S45] and Sequence to Sequence (Seq2Seq) [S45] models. A set of studies [S45, S48] showed that adding augmented data to train HIDS using representation ML models can improve HIDS performance. Besides, a study [S48] showed that GAN provides a more reliable way to mitigate data imbalance problem using data augmentation than SMOTE over-sampling technique.\\

\textbf{To summarize the NLP methods used in HIDS}, word embedding, language modeling and hybrid methods (i.e., combination of multiple NLP methods) are the prevalent NLP methods used in HIDS. These NLP methods are also showing an increasing trend over the last 4 years as demonstrated in Table \ref{tab:det_year}. The increasing amount of large-scale host security data and evolving complex attack patterns have contributed to this increasing trend of using these NLP methods that are evolved over deep learning models due to their prominent HIDS performance. Table \ref{tab:tax} presents our deep comparison analysis of the NLP methods used in HIDS, which is expected to help the HIDS researchers and developers to choose the suitable NLP methods based on their research goal and organization's requirement, respectively.  


\begin{table*}[ht]
\centering
\caption{Key factors considered in NLP-based HIDS development with their descriptions, strengths and weaknesses}
\label{tab:factors}
\footnotesize
\begin{tabularx}{\textwidth}{p{0.09\textwidth} p{0.27\textwidth} p{0.27\textwidth} p{0.27\textwidth}}
\toprule
Type & Description & Strengths & Weaknesses\\
\hline
\multicolumn{4}{c}{\textbf{\cellcolor{gray!30} {HIDS Types}}} \tabularnewline 
Anomaly (59)& Detects deviations from normal patterns as anomalies &$\bullet$ Able to detect zero-day attacks& $\bullet$ High false alarm rate\\\hline
Misuse (6)& Detects system behaviors matching the signatures present in attack library as intrusions &$\bullet$ Low false alarm rate&$\bullet$ Unable to detect zero-day attacks\\

\multicolumn{4}{c}{\textbf{\cellcolor{gray!30}{Learning Types}}} \tabularnewline  
Semi-supervised (36)& Trains only normal samples with no anomalies in a given training data set &$\bullet$ Only the normal class's labeled data is required. &$\bullet$ Suffers from high false alarm rate\\\hline
Supervised (27)& Uses labeled data to train a learning model on normal and attack data. 
&
$\bullet$ Stable performance and effective way to detect
known attacks. &$\bullet$ Requires labeled training data that is costly and time-consuming to gather\newline  $\bullet$ Difficult to detect unknown attacks \\\hline
Unsupervised (5)& Without any prior knowledge, utilize statistical models to detect anomalies. &$\bullet$ Does not require labeled training data\newline  $\bullet$ Lower computational complexity
&$\bullet$ No access to information about distinguished data patterns\\

\multicolumn{4}{c}{\textbf{\cellcolor{gray!30}{Feature extraction techniques}}} \tabularnewline  
Automated (55)&Features are extracted using automated approaches such as NLP methods (e.g., n-gram, TF-IDF and word embedding).&$\bullet$ More efficient and repeatable\newline
$\bullet$ Less dependent on domain knowledge\newline
$\bullet$ Adaptable to new datasets\newline
$\bullet$ Suitable for complex data with structural and sequential dependencies&
$\bullet$ Computationally expensive\newline
$\bullet$ Dependent on data size and quality\\\hline
Manual (5)&Features are extracted by manual data analysis and using domain knowledge.&
$\bullet$ Interpretable features \newline
$\bullet$ Suitable for homogeneous data with linear patterns&$\bullet$ Highly depends on domain knowledge\newline
$\bullet$ Takes significant time and effort\newline
$\bullet$ Error prone\newline
$\bullet$ Not scalable and dataset-dependent\\\hline
Semi-automated (5)&Features are extracted using both manual and automated techniques &
$\bullet$ Gains advantages of both automated and manual feature extraction
&$\bullet$ Needs more time than automated\newline
$\bullet$ Needs more computation than manual\\

\multicolumn{4}{c}{\cellcolor{gray!30}\textbf{{Classifiers}}} \tabularnewline 
ML: Single (38) & A single ML model is used to perform the prediction (e.g., SVM, NB) &$\bullet$ Easy implementation \newline $\bullet$ Re-training on large scale dataset is efficient & $\bullet$ Susceptible to overfitting \newline $\bullet$ May not perform as good as ensemble or DL models\\\hline
ML: Ensemble (7) & Combine predictions from two or more models (e.g., RF, XGBoost). & $\bullet$ Usually makes better predictions than single contributing model.\newline $\bullet$ Less susceptible to overfitting & $\bullet$ Longer training time than single models\\\hline
Rule or sequence matching (17) & Rules are defined (e.g., semantic ontology) or sequence matching is performed to detect attacks  (e.g., SC4ID [S16])& $\bullet$ Usually parameter-free (apart from a decision threshold)\newline $\bullet$ Interpretable
as they enable to find the location of abnormal areas in long syscall sequence. & $\bullet$ Defining the rules and the decision threshold is manual \newline $\bullet$ Requires domain expertise.\\\hline
DL: RNN (12) & RNN is a type of NN that helps to model sequential or time series data (e.g., LSTM, GRU, BiLSTM)& $\bullet$ Hold short or long-term dependencies from sequence\newline $\bullet$ Better suited for sequence-based or textual HIDS data &
$\bullet$ Due to sequential processing, usually require longer training time than CNN\\\hline
DL: CNN (12) & CNN is a type of NN that captures local features using convolution (e.g., text-CNN, FCN, TCN) &$\bullet$ Holds local and hierarchical features\newline $\bullet$ Can perform parallel processing during training, which is faster than sequential RNN&
$\bullet$ Unable to effectively capture sequential order of data that is important for handling sequential data \\\hline

\end{tabularx}
\end{table*}

\subsection{Key Factors to Consider for Developing HIDS Using NLP}\label{subsec:key_fact}
In this section, we summarize the key factors that need to be considered by the HIDS developers for developing HIDS using different NLP methods. Table \ref{tab:factors} presents the key factors considered in NLP-based HIDS development along with their descriptions, strengths and weaknesses. Our deep comparison analysis of the categories of these factors is expected to help the HIDS developers to choose the suitable category for the corresponding factors based on the organization's requirements. We describe the key factors such as i. HIDS type (Section \ref{subsubsec:hids_type}); ii. learning type (Section \ref{subsubsec:learning_type}); iii. feature extraction method (Section \ref{subsubsec:feature_extr}) and iv. classifiers (Section \ref{subsubsec:classifier}).

\subsubsection{\textbf{HIDS type }} \label{subsubsec:hids_type}
Two types of HIDS (i.e., misuse, anomaly) are developed using NLP methods in our reviewed studies. Misuse (also called signature-based) detection uses a library of known attacks' signatures and identifies system behaviors matching the signatures present in the library as intrusions. For misuse detection, NLP methods such as semantic ontology, can be used for integrating attack signatures and information from diverse sources (e.g., cyber threat intelligence data and textual sources like security blogs and bulletins). Misuse detection gains low FAR but is unable to detect zero-day attacks \citep{review_ids_19}. However, anomaly detection or behavior-based detection builds a model based on the normal behavior of system activities and detects deviations from normal patterns as anomalies, which can detect zero-day attacks but have high FAR \citep{review_ids_19}. With the continuously evolving threat landscape, attack signatures are not always available. Hence, anomaly detection is a preferred method to detect unknown zero-day attacks. NLP methods (e.g., word embedding, language modeling) help to characterize the normal behavior of the system by preserving the semantics and contextual information of the data, which makes NLP methods highly suitable for developing anomaly-based HIDS. 

\subsubsection{\textbf{Learning type}} \label{subsubsec:learning_type}
Different learning types (i.e., supervised, unsupervised, semi-supervised) are adopted for attack detection in our reviewed studies. The choice of a learning type highly depends on the availability of labeled data. The unsupervised approach finds patterns/partitions from unlabeled data. The semi-supervised approach requires a portion of data to be labeled. However, supervised learning requires complete labeled data. In the HIDS domain, normal data is highly available (generated by the normal execution of programs within a system) but malicious data is insufficient (which requires simulating the increasing attack types). Thus, semi-supervised (used by 55.4\% reviewed studies) anomaly detection by training with only normal samples and detecting the deviations from the learned model as anomalous is prevalent in the reviewed studies. NLP methods (e.g., n-gram, word embedding) help to adopt a semi-supervised approach by learning the normal behavior of the system through the semantics and context of syscall sequence. However, semi-supervised learning lead to high FAR as they classify the unseen normal behavior as attacks. Hence, diverse NLP-based data augmentation techniques (e.g., SeqGAN [S45]) are being used to balance the HIDS dataset for adopting a supervised approach that can lower the FAR.

\subsubsection{\textbf{Feature extraction techniques }} \label{subsubsec:feature_extr}
The collected raw syscall data need to be represented by appropriate features for training the HIDS detection models. Feature extraction techniques used in the reviewed studies are manual, automated or semi-automated (e.g., both manual and automated). Manual features are extracted by manual data analysis and using domain knowledge. However, the use of NLP methods such as n-gram (e.g. [S11, S17]), TF-IDF (e.g. [S55]) and word embedding (e.g. [S47, S51]) in HIDS enabled automated feature extraction from HIDS dataset and made it suitable for complex data with structural and sequential dependencies. Automated feature extraction is a preferred choice as it reduces human effort and can adapt to new datasets. Hence, in our reviewed studies automated feature extraction method is prevalent (used by 84.6\% reviewed studies) as it does not depend on domain expertise and makes the HIDS scalable and portable. 

\subsubsection{\textbf{Classifier }} \label{subsubsec:classifier}
The extracted NLP-based features are used in three types of classifiers (i.e., traditional ML, rule-based and DL model) for detecting intrusions in HIDS. While the earlier research in NLP-based HIDS used the traditional ML-based classifiers (e.g., Support Vector Machines (SVM), Logistic Regression (LR)), in the recent five years (i.e., 2018-2022), the trend is shifted to the application of DL-based (e.g., text-CNN) and hybrid classifiers. The possible reason behind the change in the trend can be the highly increasing amount of data and evolving complex attacks [S22]. DL is now being investigated more as it has the capability to handle large-scale data and the complex DL structure tends to provide better accuracy and lower FAR. 

\textbf{Traditional ML models:} For the traditional ML models, single models (e.g., SVM, Naïve Bayes (NB)) were prevalent in the earlier research time due to their simplicity. However, single models do not perform as good as the ensemble of DL models. Among the single models, SVM and its variants are the most popular algorithms (e.g., SVM, OCSVM) \citep{svm_ids_survey} used for binary and one-class classification adopting supervised and semi-supervised learning, respectively. The popularity of SVM is expected as SVM works well with the commonly used NLP-based methods such as BoW and TF-IDF \citep{slr_triet_vul}. For example, OCSVM with variable length n-gram features [S41] outperformed the approaches using TF/TF-IDF features or using HMM, K-Nearest Neighbour (KNN) and STIDE for attack detection. Another study [S59] proposed n-gram-based and frequency of n-gram-based approaches, where the frequency of n-gram-based OCSVM outperformed instance-based methods (e.g., KNN, K-Furthest Neighbors (KFN)) of prior studies [S28, S31], and had lower FAR than KFN. Besides SVM, other commonly used single ML classifiers used for HIDS are instance-based (e.g., KNN and KFN) and NN-based (e.g., MLP and Extreme Learning Machine (ELM)) single models. Moreover, ensemble ML-based classifiers (e.g., Random Forest (RF), EXtreme Gradient Boosting (XGBoost)) are used as they are less prone to overfitting and usually perform better than single ML-based classifiers. For example, for real-time detection of applications which invoke a large number of syscall, a study [S3] used RF. RF based on NLP-based TF-IDF of syscall sequences outperformed various single classifiers such as Decision Tree (DT), KNN, MLP, Multi-variable NB (MNB) and SVM [S3].

\textbf{Rule or sequence matching:} Another common approach adopted in NLP-based HIDS is the rule (e.g., semantic ontology) or sequence matching (e.g., SC4ID [S16]) to classify normal and anomalous HIDS input syscall sequence. For example, a study [S16] used a similarity approach SC4ID considering the minimal number of subsequences required to build a complete covering of a given sequence, which outperformed other text similarity measures (e.g., Levenshtein’s distance, Longest Common Subsequence/Substring) in terms of accuracy and execution time. The rule or sequence matching approaches are interpretable as they enable to find the location of abnormal areas in long syscall sequences. However, defining the rules and the decision threshold is manual and requires domain expertise. 

\textbf{DL models:} The adoption of NLP-based DL models gained tremendous popularity in HIDS from 2018 and showed better results than traditional ML and rule or sequence matching-based models. In the HIDS domain, the use of DL models not only enabled anomaly detection (binary classification) and attack detection (multi-class classification for specific attack detection) but also enabled prediction of the future syscall during an attack. The prevalent practice in NLP-based HIDS is to use variants of RNN (e.g., [S11, S12, S38, S43, S47, S51, S62]) and CNN (e.g., [S8 , S9, S17, S22, S48, S57, S63]), while a few studies used other DL models such as Deep Multi-layer Perceptron (Deep MLP) [S18, S53], Autoencoder [S9, S38, S43, S62] and Deep Belief Network (DBN) [S55]. Due to RNN's \citep{dl_survey} intrinsic capability of handling sequence data that can capture short or long-term dependencies of the input sequence, a wide variety of RNN (e.g., LSTM, GRU, BiLSTM, CuDNNLSTM) are frequently used in syscall sequence-based HIDS. However, RNN works in a sequential manner and takes longer training time than CNN as the CNN model can be trained in parallel. Different CNN models (e.g., text-CNN, Fully Convolutional Network (FCN), Temporal Convolutional Neural Network (TCN)) have been adopted in HIDS. While CNN can capture the local and hierarchical features, it can not effectively capture the sequential order of data which is required for syscall sequence data handling. The latest trend is to ensemble CNN and RNN models to gain the advantage of both models, where CNN extracts local syscall relation features and RNN captures syscall sequences context by extracting long-distance sequence dependency. For example, a study [S4] used 7 ensemble DL classifiers such as dual-flow (LSTM-FCN, GRU-FCN), FCN and Windows OS data-specific (i.e., AWSCTD dataset) models (CNN-Dynamic (CNN-D), CNN-LSTM, CNN-GRU and CNN-Static (CNN-S)). However, the complex dual-flow models did not show any advantage over the single-flow models (e.g., CNN-S) for data processing and increased training and detection time, which hinders real-time detection [S4].\\

\textbf{To summarize the key factors considered while developing NLP-based HIDS}, the prevalent practice in terms of HIDS types is anomaly-based detection which helps to detect unknown zero-day attacks. Besides, the dominant practice in terms of learning types used for NLP-based HIDS is semi-supervised learning approach which helps to address the lack of balanced available HIDS datasets. In addition, the dominant practice in terms of feature extraction used for NLP-based HIDS is automated feature extraction which helps to extract features from complex HIDS data with structural and sequential dependencies. The adoption of NLP-based DL models helped to achieve better results than traditional ML and rule or sequence matching-based models for HIDS using NLP methods, which gained tremendous popularity in HIDS since 2018 in terms of classifiers used for NLP-based HIDS.

\section{RQ2: Evaluation of NLP Method-based HIDS}\label{sec:RQ2}
To answer RQ2, we analyze the evaluation of NLP method-based HIDS in terms of 

$\bullet$ RQ2.1 Attacks detected by NLP method-based HIDS (Section \ref{subsec:attacks});

$\bullet$ RQ2.2 Datasets used to apply NLP methods to develop HIDS (in Section \ref{subsec:dataset}); and 

$\bullet$ RQ2.3 Evaluation metrics that are used to evaluate NLP method-based HIDS (in Section \ref{subsec:metrics}).


\begin{table*}[ht]
\centering
\caption{Categorization of attacks identified in NLP-based HIDS} \label{tab:attack_cat}
\footnotesize
\begin{tabularx}{\textwidth}{p{0.07\textwidth} p{0.1\textwidth} p{0.34\textwidth} p{0.41\textwidth}}
\toprule
Attack& Impacted & Attack Instances & Study Ref\\
Type& Requirement &  & \\\hline
\midrule
R2L (55)&	Integrity & imap, xlock, sshtrojan, ppmacro, netbus, sendmail, snmpget, ncftp, httptunnel, xsnoop, named, dict, framespoof, netcat, guest, ftpwrite, phf, WebShell, ShellCode, Tomcat v6.0.20, decode & S1, S2, S3, S5, S6, S7, S8, S9, S11, S12, S14, S15, S16, S17, S18, S19, S20, S21, S22, S23, S24, S25, S27, S28, S29, S30, S31, S33, S35, S37, S38, S40, S41, S42, S43, S44, S45, S46, S47, S48, S49, S50, S51, S52, S53, S54, S55, S56, S57, S58, S59, S60, S61, S62, S63\\ \hline
U2R (53) & Integrity &	Adduser, sechole, ps, secret, perl, fdformat, casesen, ntfsdos, yaga, ppmacro, eject, loadmodule, nukepw, sqlattack, xterm, ffbconfig, sunsendmailcp, syslog, CVE-2015-5602, CVE-2016-5195 & S1, S2, S3, S5, S6, S7, S8, S9, S11, S14, S15, S16, S17, S18, S19, S20, S21, S22, S23, S24, S25, S27, S28, S29, S30, S31, S35, S37, S38, S40, S41, S42, S43, S44, S45, S46, S47, S48, S49, S50, S51, S52, S53, S54, S55, S56, S57, S58, S59, S60, S61, S62, S63\\ \hline
ACE (48)& All
& Meterpreter, Java-Meterpreter, XAMPP Lite v1.7.3, Icecast v2.0, OS SMB, OS Print Spool, PMWiki v2.2.30, Wireless Karma, Adobe Reader 9.3.0, IE v6.0.2900.2180, Infectious Media, CVE 2012-0911, CVE-2009- 0927, PDFKA, CVE-2014-3120, CVE-2015-1427, CVE-2014-6271, Zip Slip, Redis, PHP-FPM & S1, S2, S3, S5, S6, S7, S8, S9, S11, S12, S14, S15, S16, S17, S18, S19, S20, S21, S22, S23, S24, S25, S28, S29, S30, S31, S33, S35, S36, S37, S38, S40, S41, S42, S43, S44, S45, S46, S47, S48, S50, S51, S54, S55, S57, S59, S62, S63\\\hline
Brute force (45)& Confidentiality &  Password bruteforce (Hydra, Hydra-FTP, Hydra-SSH), CVE 2012-2122, CWE-307 & S1, S2, S3, S5, S6, S7, S8, S9, S11, S14, S15, S16, S17, S18, S19, S20, S21, S22, S23, S24, S25, S28, S29, S30, S31, S35, S37, S38, S40, S41, S42, S43, S44, S45, S46, S47, S48, S50, S51, S54, S55, S57, S59, S62, S63\\ \hline
DoS (21)& Availability &selfping, dosnuke, back, tcpreset, syslogd, arppoison, mailbomb, teardrop, processtable, neptune, udpstorm, land, warezclient, apache2, crashiis, smurf, pod, DDoS, CVE 2012-0021, CesarFTP 0.99g, forwarding loops, CVE-2016-6515&	S1, S6, S7, S9, S12, S16, S17, S18, S23, S24, S27, S30, S42, S49, S52, S53, S54, S56, S58, S60, S61\\ \hline
Backdoors (10)&	Confidentiality, \newline Integrity & Backdoor executables , Win32.Hydraq & S1, S4, S6,  S9, S10, S12, S17, S36, S42, S54\\\hline
Worm (6)& Integrity, \newline Availability & W32.Deborm.Y, W32.HLLW.Doomjuice.B, W32.Korgo.X, W32.Sasser.D, Daber.A, Slackor.A, NetSky.y, Mytob.x, Zhelatin.uq & S1, S9, S17, S35, S36, S64\\ \hline
Trojan (5)& All
& Ransom, Downloader, Spy, PSW, Dropper, Clicker, AdWare, WebToolbar, DangerousObject, RiskTool, Wannacry & S4, S10, S26, S35, S64\\ \hline
Data Theft (4)&Confidentiality& Reconnaissance, CVE 2011-1153, CVE4-2014-0160 &	S1, S9, S17, S24\\ \hline
Probe (3)&	Confidentiality& portsweep, queso, msscan, lsdomain, illegal-snifer, ipsweep ntinfoscan, satan, doppelganger, chimera, chameleon &	S12, S18, S30\\ \hline
Virus (3)& All
& W32/Virut.n, Z0mbie.MistFall.3 & S4, S10, S36\\ \hline
Misc (25)& All
& lprcp, sm565a,  sm5x, Packed, generic, IE deleted object vulnerability, CWE5-434, CWE-89 & S1, S4, S7, S9, S10, S13, S16, S17, S18, S23, S27, S30, S32, S34, S36, S38, S39, S49, S52, S53, S56, S58, S60, S61, S65\\\hline
\end{tabularx}
\end{table*}

\subsection{RQ2.1: Attacks Detected by NLP Method-based HIDS}\label{subsec:attacks}
Our motivation for studying and categorizing security attacks is to contextualize the attacks that are targeted to be detected by the reviewed NLP-based HIDS. To answer RQ2, we categorize attacks that are detected using NLP methods in HIDS by the reviewed papers, where the HIDS is evaluated against a dataset including particular attacks. 133 instances were reported as attacks in the reviewed studies, which we categorized into 12 attack categories. Table \ref{tab:attack_cat} shows the attack categories and highlights the impacts of the attacks on the security requirements, which provide a deep understanding of security attacks and the invasion target. The attack categorization was adapted and adjusted from a prior study \citep{review_ids_21}. Later, we show the mapping of the attack categories adopted in the used datasets for NLP-based HIDS in Figure \ref{fig:ds_attack} (b) of Section \ref{subsec:dataset}. 

\subsubsection{\textbf{Attack impacts on security requirements}} 
We consider the three most significant security requirements (i.e., Confidentiality-Integrity-Availability (CIA)), that need to be provided by a security framework \citep{c3i}. These requirements are impacted by security attacks, and can leave disastrous consequences on host systems, such as loss of personal information and business reputation along with financial losses due to the impact on the requirements by security attacks. These requirements are also known as CIA triad, which we describe as follows. 

\begin{itemize}[noitemsep,topsep=0pt]
    \item \textbf{Confidentiality} ensures data security by preventing information from being disclosed to unauthorized individuals, entities or processes. 
    \item \textbf{Data integrity} protects data from unauthorized modifications throughout its life cycle, ensuring its accuracy, trustworthiness and validity.
    \item \textbf{Availability} ensures the availability of information or services for legitimate users upon demand.
\end{itemize}

\subsubsection{\textbf{Categorization of attacks detected in NLP-based HIDS}} In this section, we discuss the attack categories that can be detected using NLP-based HIDS, in terms of included attacks and impacts on the above security requirements. 


\textbf{\textit{(1) Remote to Local (R2L)}} is the most reported attack type in our reviewed studies. It involves attackers exploiting the host’s vulnerability to gain illegal local access and affects data integrity \citep{thomas2008usefulness}. An example of R2L attack instance is \textit{Tomcat v6.0.20}, which enables remote intruders to execute unrestricted file upload attacks. Another example is \textit{ncftp}, which is a Linux FTP file access-utility version with a bug that allows remote commands to execute on a local host.

\textbf{\textit{(2) User to Root (U2R)}} 
attack enables intruders to gain a system's root access starting with access to a normal user account (achieved by password sniffing, dictionary attack, or social engineering) \citep{thomas2008usefulness}. It bypasses the authentication and threatens the data integrity by removing security policy-specified files from the victim hosts. An example of U2R attack instance is \textit{Adduser} that creates new superuser to gain root access to a host utilizing a corrupted executable. 
\textbf{\textit{(3) Arbitrary Code Execution (ACE)}} involves an attacker gaining control by injecting his own code by exploiting some vulnerability. It can affect any of the security requirements depending on the target of the executed arbitrary code/commands on a target host \citep{simmons2014avoidit}. An example of ACE attack instance is \textit{XAMPP Lite v1.7.3} that permits a remote intruder to run malicious payload utilizing Xampp\_webdav application. Another ACE example is \textit{Icecast v2.0} that allows remote attackers executing arbitrary code using an HTTP request.


\textbf{\textit{(4) Brute force}} 
attack employs a trial and error process that generates a huge number of guesses and validates them to collect data (e.g., account password, personal id number), which damages the data confidentiality and can circumvent the authentication by acquiring an authorized user login information. For example, a vulnerability (\textit{CVE 2012-2122}) of MySQL 14.14 allows intruders to bypass the authentication process by authenticating the same incorrect password repeatedly. 


\textbf{\textit{(5) Denial of Service (DoS)}} 
attack denies the legitimate user access to a host machine or service by making the computation resources extremely busy to manage legitimate requests \citep{simmons2014avoidit}. Examples of DoS include exploiting Apache web server vulnerability \textit{CVE 2012-0021} for causing a daemon crash using an empty cookie [S24] and \textit{tcpreset} that terminates host TCP connections \citep{thomas2008usefulness}. Distributed DoS (DDoS) attack uses multiple machines to flood a targeted resource that is detected in a few reviewed studies [S1, S17]. Besides, forwarding loops are potential DoS attacks as they cause one request to be processed repeatedly or even indefinitely, resulting in undesired resource consumption that is detected in several reviewed studies (e.g., [S16, S52]). 

\textbf{\textit{(6) Backdoor}} secures remote access to a machine, or plain text in a cryptographic system, which attackers can use to get access to sensitive host data (e.g., passwords), alter or remove information on hard drives and transfer data across networks. Backdoor attack impacts confidentiality and integrity. For example, \textit{Win32.Hydraqa} is a family of backdoors [S36]. 

\textbf{\textit{(7) Worm}} is a malware type, which replicates itself to spread to uninfected hosts throughout the network without human intervention \citep{simmons2014avoidit}. Worms can damage service availability by consuming bandwidth/storage space, and affect data integrity by corrupting or modifying files on a target host. For example, \textit{NetSky.y}, \textit{Zhelatin.uq} and \textit{Mytob.x} are worms, which can overwrite other executables and try to exploit host OS components [S36]. 

\textbf{\textit{(8) Trojan}} is a program that appears appealing and legitimate but has anomalous code in it, which can affect any of the security requirements. For example, \textit{AdWare} is a software which automatically shows advertisement pop-ups while a user is online. Other examples include \textit{Trojan-PSW} which steals user account information, \textit{WebToolbar} that installs in-browser content without users’ consent and Trojan-Ransom that prevents user's access to demand payment. 

\textbf{\textit{(9) Data theft}} threatens data confidentiality by stealing information stored on corporate databases, devices and servers. An instance of data theft is observed in a study [S24], which uses PHP 5.3.5 vulnerability \textit{CVE 2011-1153} that steals context-sensitive data from the memory of the host process. 

\textbf{\textit{(10) Probe}} attacks scan a host automatically to gather records of private systems or a DNS server that devastates data confidentiality \citep{thomas2008usefulness}. For example, \textit{ipsweep} and \textit{lsdomain} enable finding legitimate IP addresses. \textit{Mscan} and \textit{queso} enable finding host OS types. \textit{Portsweep} enables finding active ports. 



\textbf{\textit{(11) Virus}} is a code that attaches itself through any infected host files and self-replicates when the program is run \citep{simmons2014avoidit}. Based on the target of the executed code, a virus can pose a threat to each security requirement. Examples include \textit{W32/Virut.n} which is a polymorphic virus infecting HTML files and \textit{Z0mbie.MistFall.3} metamorphic virus infecting other executables [S36]. 

   
\textbf{\textit{(12) Miscellaneous}} category covers the remaining types of attacks that can affect different security requirements depending on the target of performed activities. If the details of the detected attack (i.e., anomaly) are not presented in the corresponding reviewed study, we referred them under this category. Examples include \textit{lprcp} attack script that replaces arbitrary file contents and \textit{Packed} that refers to compressed obfuscated malicious programs, which cannot be analyzed. Another example is exploiting Internet Explorer's (IE) deleted object vulnerability [S36]. \\

\begin{figure*}[]
  \centering
  \includegraphics[width=\textwidth]{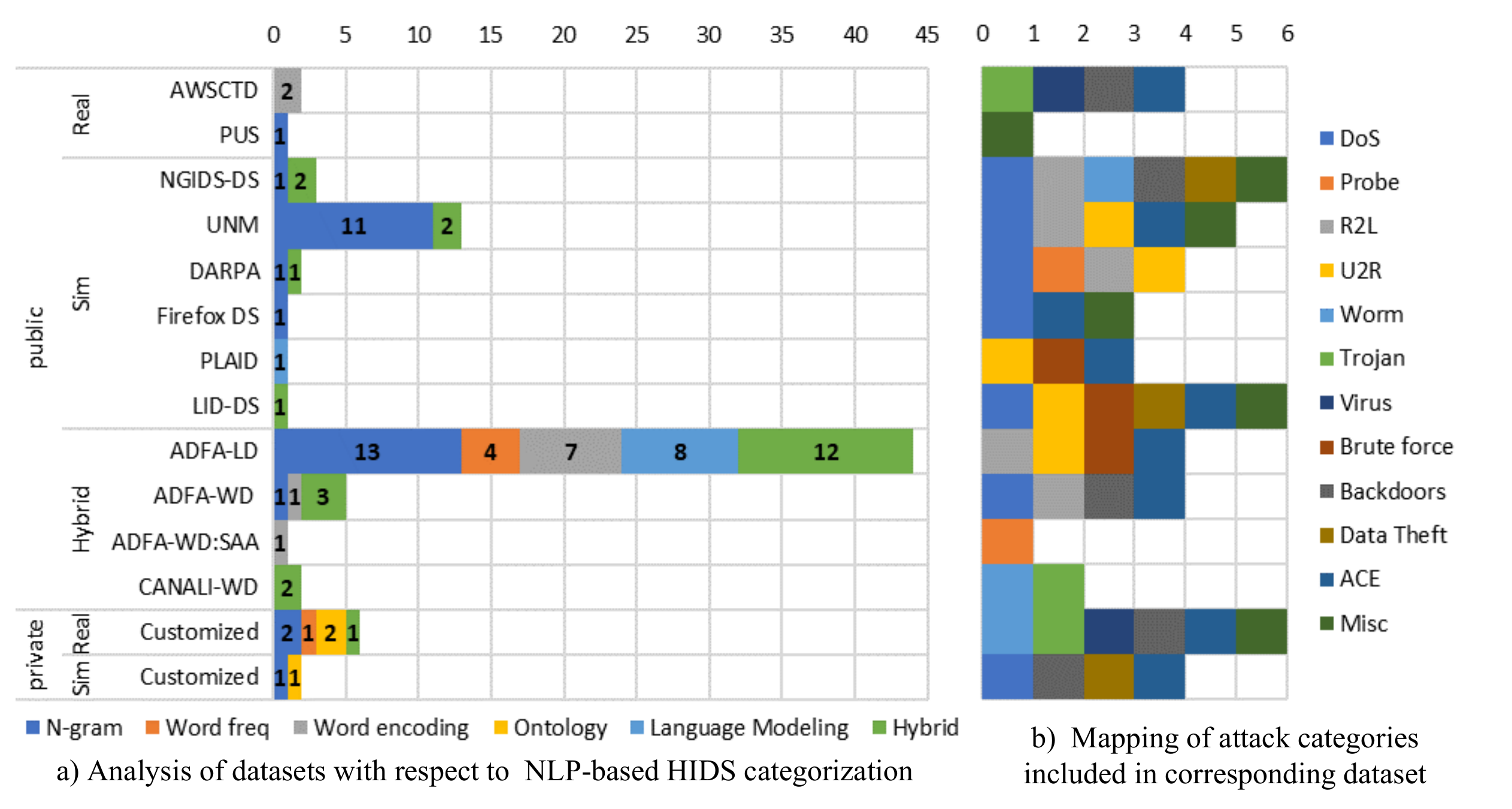}
  \caption{Analysis of datasets grouped by type and availability with respect to NLP solutions categories and attack types }
  \label{fig:ds_attack}
  
\end{figure*}

{\footnotesize
\begin{table*}[ht]
\centering
\caption{Type of NLP-based HIDS datasets, their description with strengths and weaknesses}
\label{tab:ds_types}
\footnotesize
\begin{tabularx}{\textwidth}{lXXX}
\toprule
Type & Description &Strengths & Weaknesses\\
\midrule
Real & Datasets including real data captured from a real organization/production environment. Both data and environment are real. & $\bullet$ Providing true distribution of data& $\bullet$ Imbalanced datasets with an insufficient number of malicious activity\newline
$\bullet$ Covering a limited attack types
\\\hline
Simulated& Dataset including either synthetic data (e.g., artificially generated data) or data captured within a test bed or emulated controlled environment. & $\bullet$ Able to reproduce balanced
datasets.\newline
$\bullet$ Able to generate rare misuse events.\newline
$\bullet$ Useful for attacks for which real data is not available.& $\bullet$ Tool specific\newline
$\bullet$ May not depict the real distribution of data\newline
$\bullet$ May not represent a real heterogeneous
environment.\\\hline
\end{tabularx}
\end{table*}
}

\textbf{To summarize RQ2.1}, we identified that all the NLP method categories (described in Section \ref{subsec:nlp_cat}) have been explored to detect the top 3 frequent attack types such as R2L, U2R and ACE, as shown in Table \ref{tab:attack_cat}. Besides, hybrid NLP-methods have been explored to detect all attack categories. Semantic ontology has the ability to correlate multiple attack instances, which enable to dynamically detect diverse attacks in HIDS [S13] and language modeling has been shown to significantly improve the HIDS detection performance [S5], however, we observed ontology and language modeling are yet to explore the detection of the different attack categories (e.g., DoS, Probe, Virus and Backdoor). We present Table \ref{tab:attack_cat} to help researchers and developers to learn about the attack types, the relevant attack instances and impact of these attacks that they aim to detect using NLP-based HIDS.


\subsection{RQ2.2: Datasets Used to Apply NLP Methods for Developing HIDS } \label{subsec:dataset}
We identified 20 different datasets employed in the reviewed NLP-based HIDS. We discuss the characteristics of these datasets with respect to types of datasets, availability, included attack and NLP methods used on these datasets. Figure \ref{fig:ds_attack} shows the distribution of the datasets along with their types of datasets and availability with respect to NLP-based HIDS solutions categorization. Figure \ref{fig:ds_attack} also shows the different attack categories that are present in the corresponding dataset along with the total attack categories available in that dataset to show the diversity of the included attacks. Figure \ref{fig:ds_attack} presents a comprehensive overview of the datasets to help practitioners and researchers to understand these characteristics of the datasets.

The primary data source of these datasets that are used to apply NLP methods in HIDS research is syscall sequence. Syscall sequence is a primary artifact of the OS kernel. Syscall sequence is considered the most reliable data source for intrusion detection as they represent low-level information without any filtering and processing [S4]. Since in syscall sequence, any filtering, interpretation and processing are not used, it can not obfuscate events [S40]. Datasets have been generated by collecting syscall sequences either in real (i.e., organization/production environment), simulated (either synthetic data or controlled environment (e.g., testbed, emulation environment, lab)) or hybrid environments. Inspired by an existing dataset classification \citep{dataset_survey}, we have categorized datasets used by the HIDS researchers into real, simulated and hybrid categories to identify and characterize the existing practices adopted to evaluate HIDS. While real dataset represents the true distribution of data, it usually provides imbalanced dataset with limited attack types. In contrast, simulated dataset is able to generate balanced dataset, which may not depict the true data distribution. Hybrid dataset can gain the advantage of both real and simulated dataset. Table \ref{tab:ds_types} describes the types of datasets along with their strengths and weaknesses. The details of the study mapping with the dataset types are available in Table A.2 of our online appendix \citep{online_appendix}.

The availability of the identified datasets is categorized as public and private (i.e., customized). Public datasets in the HIDS domain are usually outdated, lack sufficient labeled data and do not cover a wide variety of attack types as shown in Figure \ref{fig:ds_attack}. To overcome these limitations, some studies were motivated to explore methods for generating new customized datasets, which are usually kept private. Table \ref{tab:public_ds} presents a list of currently accessible public datasets used in our reviewed studies with their description, strengths and weaknesses to help HIDS researchers and developers easily identify and choose the required public datasets to develop or evaluate their NLP-based HIDS. 

We discuss the identified datasets in terms of types and availability such as we discuss public simulated dataset in Section \ref{subsubsec:pub_sim}, public real dataset in Section \ref{subsubsec:pub_real}, public hybrid dataset in Section \ref{subsubsec:pub_hyb}, private simulate dataset in Section \ref{subsubsec:pri_sim} and private real dataset in Section \ref{subsubsec:pri_real}.




\subsubsection{\textbf{Public simulated dataset}}\label{subsubsec:pub_sim}
We identified six public simulated datasets such as i. Defence Advanced Research Project Agency (DARPA) or Knowledge Discovery and Data Mining (KDD); ii. University of New Mexico (UNM); iii. Firefox DS; iv. Next Generation Intrusion Detection Systems Data Set (NGIDS-DS); v. Leipzig Intrusion Detection-Data Set (LID-DS) and vi. Lab Artificial Intrusion Dataset (PLAID), which we discuss in this section. 

\begin{table*}[]
\centering
\caption{Public datasets used in reviewed NLP-based HIDS with description, strengths and weaknesses}
\label{tab:public_ds}
\footnotesize
\begin{tabularx}{\textwidth}{p{1.78cm}XXX}
\toprule
Dataset & Description & Strengths & Weaknesses\\
\midrule
DARPA/KDD \citep{cunningham1999darpa} \newline1998/99& Includes Basic Security Module (BSM) data file with syscall-based audit data produced in a victim’s machine for host-level audit. & $\bullet$ First standard corpora for evaluation of NIDS and widely used as benchmark\newline $\bullet$ Includes arguments and return values& $\bullet$ Very obsolete, unable to accommodate the latest trend in attacks\newline
$\bullet$ Focus on NIDS and lacks the information required to train HIDS-suitable methods\\\hline

PUS-DS [S58] \newline 2000 & A syscall dataset collected from 8 users within 2 years. One of the user’s data is considered normal data, and
it is mixed with few other user data to consider abnormal & $\bullet$ Real dataset collected from eight different users & $\bullet$ Lacks attack scenarios and attack diversity \newline $\bullet$ Outdated, unable to represent latest user behavior \\\hline

UNM \citep{unm} \newline 2004& Includes synthetic sendmail UNM, synthetic sendmail CERT, live lpr UNM and live lpr MIT datasets. Synthetic traces were collected by running a prepared script. & $\bullet$ Includes programs of varied size and complexity, and different kinds of intrusions (buffer overflows, symbolic link attacks and Trojan programs)& $\bullet$ Very obsolete\newline $\bullet$ Lacks syscall arguments or other metadata\newline $\bullet$ Extremely limited in scope and not represent a full sampling of OS, focus on single processes (process IDs, syscall IDs)\\\hline
CANALI-WD \citep{canali2012quantitative}\newline 2012& Includes program execution traces observed both in a synthetic environment and on real-world machines with actual users and under normal operating conditions.   & $\bullet$ Presents a large collection of anomalous traces compared to previously published syscall datasets (e.g., UNM).\newline $\bullet$ Not biased towards particular runtime environments, or usage patterns.&$\bullet$ Lacks some useful information such as syscall arguments, timestamp, etc. \\\hline

Firefox DS \citep{firefox_ds} \newline 2013 & Includes normal traces of Firefox 3.5 by executing 7 testing frameworks and anomalous traces by launching contemporary attacks against Firefox. & $\bullet$ The completeness of Firefox's normal behavior is ensured using code coverage for test-case execution, which resulted in 60\% source code coverage & $\bullet$ Lacks attack diversity \newline $\bullet$ All the traces are specific to firefox.\\\hline
ADFA-LD \citep{creech2013adfa} \newline 2013 & Alternative to older datasets (DARPA, UNM) and collected under Ubuntu OS running services and simulating attacks.& $\bullet$ To attain a realistic defensive and attacking environment, a system was attacked by a certified penetration tester with 60 different attacks belonging to U2R, R2L, ACE and brute force attack types.
&$\bullet$ Outdated and not representative of contemporary attacks\newline $\bullet$ Lacks syscall argument, return data and other metadata\\\hline
ADFA-WD \citep{creech2013adfa} \newline 2013&  Collected on a Windows host, and a total of 12 known vulnerabilities were exploited to simulate different attack types.& $\bullet$ Real vulnerabilities in commonly used software were considered for generating attack data & $\bullet$ Outdated and not representative of contemporary attacks\newline $\bullet$ Lacks syscall arguments, return values, or other metadata\newline
$\bullet$ Inadequate number of vulnerabilities used to create malicious activity \citep{vceponis2018awsctd}\\\hline
ADFA-WD:SAA \citep{adfa_wd} 2016&An extension of ADFA-WD which includes Windows-based stealth attacks&$\bullet$ Suitable for evaluation against stealth attacks&$\bullet$ Lacks metadata\newline $\bullet$ Attack types limited to stealth attacks\\\hline
NGIDS-DS \citep{haider2017ngids} \newline 2017 & Obtained from Ubuntu 14.0.4 host that is equipped with an auditing mechanism and includes 99 host log files.&  $\bullet$ Up-to-date and synthetically realistic\newline $\bullet$ Includes thread info, timestamp, eventID, path, processID, syscall, etc.
 &$\bullet$ Loss of parameters and more accurate timestamps\\\hline
AWSCTD \citep{vceponis2018awsctd}\newline 2018 & Includes benign
software samples as benign data and public malware files from Virus Share \citep{VirusShare} and publicly available data about the malwares identified from Virus Total \citep{VirusTotal} as attack data. & $\bullet$ New extended dataset for Windows\newline $\bullet$ Includes parameters (syscall args, return value) for in-depth training & $\bullet$ Includes only different Malware types and lacks other types of diverse attacks.\newline \\\hline

LID-DS \citep{grimmer2019modern}\newline 2019 & A modern data set collected from a modern OS (Ubuntu 18.04) by considering diverse scenarios of real vulnerabilities. & $\bullet$ Consists of different real vulnerabilities scenarios.\newline $\bullet$ Includes unambiguous syscall sequence and their thread assignment of recent multi-threaded scenarios. \newline $\bullet$ Includes parameter, return data, user id, process/thread id, file system handle, timestamp and I/O buffer& $\bullet$ Only available for Linux OS\\\hline
PLAID [S21] \newline 2021 &Includes modern system calls and contemporary attack types collected on Ubuntu 18.04 LTS.&$\bullet$ The most up-to-date HIDS dataset currently available. \newline $\bullet$ Includes attacks in multi-process applications. & $\bullet$ Lacks syscall arguments, return values, or other metadata.\newline $\bullet$ Includes only six attack vectors. \\\hline
\end{tabularx}
\begin{tablenotes}
      \item\label{tnote:robots-r2} \footnotesize{*PUS-DS and Firefox-DS are not currently accessible}
      \end{tablenotes}
\end{table*}

\textbf{DARPA} or \textbf{KDD \citep{cunningham1999darpa}}: DARPA or KDD are the earliest efforts of IDS dataset which are focused on NIDS data and lack sufficient information needed to train HIDS using suitable NLP methods [S10]. The DARPA dataset was generated in a testbed and includes information of syscalls generated by running programs for 7 weeks for the training dataset and 2 weeks for the test dataset [S18]. Though DARPA includes DoS, R2L, U2R and Probe attacks, it is considered obsolete as it is unable to accommodate the latest trend in attacks \citep{moustafa2015unsw}. The hardware and the corresponding OS with their syscalls of this outdated dataset have changed over time. DARPA and KDD datasets were adopted in two studies [S18, S30] along with the UNM and Australian Defense Force Academy Linux Dataset (ADFA-LD) datasets to cover diverse complexity levels (i.e., syscall sequence generated by different running programs), where DARPA represents the simplest dataset.

\textbf{UNM \citep{unm}:} UNM includes program traces which were collected during live execution of privileged host programs. It also includes simulated traces which were collected using a script to collect `xlock' program's commands [S53]. Every trace includes a complete set of syscalls made during the program execution (e.g., login, sendmail) from start to end. A few studies [S60, S62] used syscalls of only specific programs (e.g., sendmail [S60, S61]). 
UNM dataset includes different attacks such as U2R, R2L and DoS. UNM being an old dataset, most of the studies that adopted this dataset used the classical n-gram-based NLP method with diverse classifiers. In spite of a remarkable usage of the UNM dataset in developing HIDS research, that dataset is marked as outdated and limited in scope \citep{creech2013adfa}. UNM does not reflect the sophistication
of modern attacks, and both normal and attack traces no longer represent the complexity of current host systems due to their outdatedness [S41]. Besides, the benign syscall traces in UNM dataset were generated by executing small programs for a longer time period that generated the same syscall execution paths [S23]. Still, they are being used for benchmarking HIDS [S18, S30] along with the newer datasets to show the effectiveness of diverse datasets as there is a lack of many publicly accessible datasets.

\textbf{Firefox DS \citep{firefox_ds}:} To overcome the above-mentioned issues of UNM dataset, firefox dataset was created for Firefox web browser. It includes 700 normal syscall traces of Firefox3.5 by executing 7 diverse testing frameworks, where every testing framework executed
various Firefox components and functionalities [S23]. This dataset includes 19 anomalous syscall traces by launching contemporary attacks including ACE and DoS attack types against Firefox, selected from public advisories and resources (e.g., Metasploit \citep{metasploit}). However, Firefox-DS is a small-scale dataset including a very limited number of attack traces. Hence, when using this dataset, other large-scale datasets should be used to evaluate the scalability of the proposed NLP-based HIDS.

\textbf{NGIDS-DS \citep{haider2017ngids}:} It is a relatively new Linux OS-based dataset that is generated in an emulated environment \citep{dataset_survey}. NGIDS-DS includes 99 host log files, where each record represents all the corresponding information (e.g., syscalls, path and attack category) about an occurred event, for both normal and anomalous activities (e.g., DoS, worm and backdoor). Only n-gram-based and hybrid approaches were adopted by the reviewed studies to evaluate HIDS using this dataset. Though NGIDS-DS includes diverse attacks and thread information, more accurate timestamps are missing, which makes it impossible to determine a deterministic order of syscalls reliably [S9]. Since several syscalls are with the same timestamp, and event-ids' order sometimes contradicts timestamps' order, reconstructing the correct sequence of syscalls is difficult.

\textbf{LID-DS \citep{dataset}:} To mitigate the incorrect sequence issue of NGIDS-DS dataset, LID-DS is released with the unambiguous syscall sequence and their thread assignment. The software of modern IT systems is multi-threaded, thus diverse syscalls may belong to different threads and users. The traditional HIDS does not consider the thread in which a syscall is executed, which consequently ignores the fact that software of modern IT systems is usually multi-threaded. 
Execution schedule of OS may generate diverse syscall sequences for multiple executions of the same software with the same input data.
Hence, this can negatively impact the accuracy of the HIDS which focuses on syscall subsequence learning and analyzing [S9]. Hence, LID-DS includes several features that are unavailable in the earlier datasets such as syscall, timestamp, thread id, process name, argument, return data and excerpt of data buffer from syscall traces of normal and anomalous behavior of multiple recent, multi-process and multi-threaded scenarios. Using the comparative analysis of different NLP methods (e.g., n-gram, word-embedding), a study [S9] showed that the included thread information from LID-DS enhances detection accuracy and reduces FAR.

\textbf{PLAID (2021) [S21]:} While HIDS traditionally considers syscall traces generated by a single process, modern applications often use multiple processes. Besides, one or multiple processes can be impacted by modern attacks. Hence, the latest syscall sequence-based dataset, PLAID, includes attacks in multi-process applications. PLAID is Linux OS-based with modern syscalls and modern attacks. PLAID includes syscall sequences from 6 current exploits and penetration methods along with a huge collection from benign operations.


\subsubsection{\textbf{Public real datasets}} \label{subsubsec:pub_real}
We identified two public real datasets such as i. Purdue Unix Shell (PUS) and ii. Attack-caused Windows OS syscall Traces Dataset (AWSCTD), which we discuss in this section. 

\textbf{PUS (2000) [S58]:} PUS is a real dataset, which is a syscall dataset collected from 8 users within 2 years [S58]. One of the user's data is considered normal data, and it is mixed with a few other user data to consider abnormal data. PUS lacks attack scenarios and attack diversity. Besides, PUS is outdated and unable to represent the latest user behavior.

\textbf{AWSCTD \citep{vceponis2018awsctd}:} AWSCTD is considered one of the biggest syscall collection running on Windows OS that includes 112.56 million syscalls from 12110 executable malware samples and 16.3 million syscalls from 3145 benign software samples. AWSCTD includes malware-initiated system calls which were collected from publicly available malware files from VirusShare \citep{VirusShare} and real world malware information from VirusTotal \citep{VirusTotal}. It includes different attacks such as worm, trojan and backdoor. AWSCTD is adopted by 2 studies [S4, S10] that showed word embedding-based CNN outperformed RNNs [S10] and complex dual-flow models including CNN and RNN models (e.g., LSTM-FCN and GRU-FCN) [S4]. As Windows is still a popular OS, HIDS researchers and developers who are interested in Windows OS-specific HIDS development can use the AWSCTD dataset as it is a real extended dataset compared to other Windows-based datasets [S10].

\subsubsection{\textbf{Public hybrid dataset}} \label{subsubsec:pub_hyb}
We identified four public hybrid datasets including i. CANALI Windows Datasets (CANALI-WD); and three Australian Defence Force Academy (ADFA) family datasets such as ii. ADFA Linux Dataset (ADFA-LD); iii. ADFA Windows Dataset (ADFA-WD) and iv. ADFA Windows Dataset: Stealth Attacks Addendum (ADFA-WD:SAA), which we discuss in this section. 

\textbf{CANALI-WD \citep{canali2012quantitative}:} It is a Windows OS-based dataset that includes program execution syscall traces observed both in simulated (based on Anubis \citep{anubis}) and the real-world environment with actual users. CANALI-WD dataset includes 2 normal datasets, namely, Goodware and Anubis-good [S64]. Goodware includes 180 GB syscall traces gathered from 10 real world host systems, used by regular users and Anubis-good includes syscall traces of 36 normal applications executed under Anubis. CANALI-WD also includes 2 malware datasets, namely, malware and malware-test. The malware dataset includes 5,855 malware (e.g., botnets and worms) syscall traces randomly extracted from Anubis and the malware-test dataset includes 1,200 syscall traces of malware collected from another machine that was not used by Anubis. While the previously published syscall datasets (e.g., DARPA and UNM) have limited malicious traces, CANALI-WD presents a huge collection of malicious traces.

\textbf{ADFA-LD \citep{creech2013adfa}:} The outdated datasets such as DARPA and UNM, which are based on more than 20 years old software which are usually no longer in use to detect or reflect the modern intricate security attacks. To replace these outdated datasets and represent modern attack structure and methodology, Australian Centre of Cyber-Security released two datasets for two different OS (i.e., Linux and Windows) named ADFA-LD and ADFA-WD, respectively. ADFA-LD is a dataset of syscall traces of Linux OS, which is the most used dataset to evaluate NLP method-based HIDS. ADFA-LD consists of syscall traces of different active services during normal host operation, where the included activities range from web browsing to LATEX document preparation. To attain a realistic defensive and attacking environment, a system was attacked by a certified penetration tester with 60 different attacks belonging to U2R, R2L, ACE and brute force attack types. During the execution of attacks, the corresponding syscall traces were recorded. ADFA-LD includes 833 normal training syscall sequences, 746 attacks and 4372 normal validation syscall sequences to evaluate HIDS. The scale of ADFA-LD dataset is considered suitable for training deep neural networks [S5, S15]. All the reported NLP categories except the semantic ontology were adopted to detect intrusion in ADFA-LD dataset by the reviewed studies. 

\textbf{ADFA-WD and ADFA-WD:SAA \citep{adfa_wd}:} These datasets are based on Windows OS, which contain syscall traces from Windows XP SP2 system. While ADFA-WD includes zero-day attacks based on vulnerabilities, ADFA-WD:SAA is created to evaluate the effectiveness of HIDS against stealth attacks. In ADFA-WD, automated hacking tools were used to exploit 12 known vulnerabilities and the attack vectors include browser attack and malware attachment. ADFA-WD includes 355 normal training syscall sequences, 5542 attacks and 1827 normal validation syscall sequences. In contrast, ADFA-WD:SAA includes 862 attack syscall traces. Since both of these syscall-based HIDS datasets are the sequence of tokens, NLP methods are well-suited to handle these datasets [S12]. 




\subsubsection{\textbf{Private simulated datasets}}\label{subsubsec:pri_sim}
In this section, we discuss the simulated datasets, which are not publicly available. To create private simulated datasets, researchers have used various libraries or tools (e.g., \textbf{Strace} [S24], \textbf{LTTng} [S24]), which allow monitoring the execution of a program and read syscall traces on user or kernel space. \textbf{Ptrace} helps to trace Linux syscall. Besides, \textit{Strace} open-source application utilizes \textit{ptrace} to provide statistics about a trace in text format. Linux Trace Toolkit Next Generation (\textit{LTTng}) tracer saves traces in Common Trace Format (CTF) \citep{desnoyers2006lttng}. Further, \textbf{drstrace} helps to trace syscall for Windows OS. A study [S10], used \textit{drstrace} to append syscall with AWSCTD dataset.

A study [S24] created a dataset based on a local Linux OS, where normal syscall traces were collected by several tasks (e.g., web browsing and document processing). For collecting attack syscall traces, Metasploit \citep{metasploit} open source penetration testing tool was utilized to exploit the popular vulnerabilities for executing the attacks. Another study [S33] created a dataset by simulating attacks in a controlled environment on a private Ethernet-based local network (includes 2 desktops and an IBM ES750 Network Scanner). This dataset enabled the study to evaluate the reasoning logic of semantic ontology on multiple diverse vulnerabilities. The reasoning logics of the semantic ontology were used to analyze the text description of the vulnerabilities and then use them to detect the possibility of attacks in logs, which successfully inferred 7120 of the 8070 attacks.

\subsubsection{\textbf{Private real datasets}} \label{subsubsec:pri_real}
A study [S32] created a dataset by collecting the normal samples from a corporate network border's live feeds of all the files, which were filtered using anti-virus scanners to ensure them as benign data. Attack data was collected from the daily feed of a security company which gathers malicious software from their network sensors. Another study [S36] gathered various real-world exploits and legitimate applications and confronted them on a Windows 7 host. The studies [S13, S26] that intend to perform knowledge fusion by semantic ontology built their own dataset by collecting data from varied dynamic textual sources (e.g., Symantec’s website \citep{Symantec}) and combining them with malware behavioral information to detect attacks. While a study [S39] collected syscall sequences on the process of sending and receiving e-mails, another study [S34] focused on Gzip software (a file compression and decompression tool for Linux) to collect data. \\

\textbf{To summarize RQ2.2}, we identified ADFA-LD \citep{creech2013adfa}, UNM \citep{unm} and ADFA-WD \citep{creech2013adfa}, are widely adopted to develop and evaluate NLP-based HIDS in practice. However, these widely adopted datasets are old to represent modern complex attacks and have only basic information (e.g., syscall identification and called function names), which is minimal to detect intricate and modern security attacks in the host [S10]. Besides, an inadequate number of vulnerabilities were utilized to generate attacks [S10] in these datasets. To reflect the modern host systems’ behavior and attacks in the host, our SLR identified several recent and extended datasets, such as AWSCTD \citep{vceponis2018awsctd} dataset for Windows OS, and NGIDS-DS \citep{haider2017ngids}, LID-DS \citep{grimmer2019modern} and PLAID [S21] datasets for Linux OS. Moreover, we observed several datasets used to develop and evaluate NLP-based HIDS in practice are private, which can hinder the reproducibility and replicability of the NLP-based HIDS. 

\subsection{RQ2.3: Evaluation Metrics}\label{subsec:metrics}
We identified 17 evaluation metrics that have been used to evaluate the performance of NLP-based HIDS. Several reviewed studies perform an intermediary step (i.e., syscall sequence prediction) and then perform the intrusion detection to enhance the detection capabilities. These studies evaluate both steps with distinguished evaluation metrics. Besides, the computation performance (efficiency) is also measured along with the detection performance (i.e., effectiveness). Hence, we categorized the evaluation metrics into three categories such as detection performance (Section \ref{subsubsec:detect_perf}), computation performance (Section \ref{subsubsec:comp_perf}) and intermediary task performance (Section \ref{subsubsec:intm_perf}). Table \ref{tab:metrics} presents the metrics of each category with their description, mathematical representation and corresponding strengths and weaknesses. This comparative analysis of Table \ref{tab:metrics} will help HIDS researchers to choose suitable metrics to evaluate the proposed HIDS. Besides, Table \ref{tab:metrics} will help the HIDS developers to reflect the organization's requirements and preferences by evaluating the HIDS based on the relevant metrics. The study mapping with the metrics is available in Table A.3 of our online appendix \citep{online_appendix}.

{\footnotesize
\begin{table*}[]
\centering
\caption{Metrics used for evaluation of NLP-based HIDS and their description, equation, strengths and weaknesses}
\label{tab:metrics}
\footnotesize
\begin{tabularx}{\textwidth}{p{0.09\textwidth} p{0.22\textwidth} p{0.11\textwidth}p{0.03\textwidth} p{0.27\textwidth} p{0.15\textwidth}}
\toprule
 Evaluation & &&Paper& &\\
Metric & Description & Equation & Count &Strengths&Weaknesses \\

\multicolumn{6}{c}{\cellcolor{gray!30}\textbf{Detection Performance (Effectiveness)}} \tabularnewline
Detection Rate (Recall, detection accuracy, TPR) & Ratio of correctly identified attacks out of total attack samples. & $\frac{TP}{TP+FN}$ & 45 &Prioritizes detecting all the attacks even at a cost of false alarm. Considering it is less costly to process all the alerts compared to missing to detect an attack, recall can be helpful. & Do not consider TN (Use of MCC can resolve this).\\\cline{1-5}
Precision &	Ratio of the correctly identified attacks to total samples identified as attacks. & $\frac{TP}{TP+FP}$&13 & Considering that handling false alarms is costly, optimized precision reveals that all the predicted alarms are worth noticing. &
\\\cline{1-5}
F-Measure &	Harmonic mean of precision and recall.&	$\frac{2\times P\times R}{P+R}$& 15 & Combines precision and recall. Suitable for highly imbalanced HIDS dataset. & \\\hline
False Alarm Rate (FAR, false positive rate, FPR)  & Ratio of incorrectly detected normal samples out of total normal samples. &$\frac{FP}{FP+TN}$& 40 & Given that the cost of dealing with a huge volume of alerts of a HIDS is high, it is an important metric for HIDS.& {Should be used as an auxiliary metric with other metrics.}
\\\cline{1-5}
False Negative Rate (Missing Rate, FNR) & Ratio of incorrectly detected attacks out of total attack samples.& $\frac{FN}{FN+TP}$ & 8 & Given that the cost of the inability to detect an attack by the HIDS model is high, it presents the missed attacks.&
\\\cline{1-5}
True Negative Rate (TNR)& Ratio of correctly identified normal samples out of total normal samples.&$\frac{TN}{TN+FP}$& 4 & To represent that the HIDS is usually correct to predict a benign behavior, it can be used.&
\\\hline
Confusion matrix & Matrix of prediction results. &-&	3 & Presents nominal value rather than normalized value to show how the HIDS model performs for different imbalance classes. Gives insight into the type of error made by the HIDS. & Rather than using it as a metric, the matrix is usually used to calculate other metrics (e.g., precision).
\\\hline
AUC based on ROC Curve (ROC-AUC)  & ROC curve shows the TPR as a function of FPR as the discrimination threshold of the classifier is varied, and AUC is the area under ROC. &- &33& Not dependent on prediction threshold. & Not suitable for heavily imbalanced HIDS data (Use of Precision-recall-AUC can resolve this).
\\\hline
Classification Acc/Rate&
Percentage of correctly classified samples. &	$\frac{TP+TN}{TP+TN+FP+FN}$	& 18 & Consider all the cells of a confusion matrix. & Not suitable for imbalanced data.
\\\hline
Classification Error &	Percentage of incorrectly classified samples. &	$\frac{FP+FN}{TP+TN+FP+FN}$	&	1 & {Suitable to represent the error of the HIDS model as the HIDS output is discrete.} & {Can be used as an auxiliary metric with other metrics.}
\\\cline{1-4}
Mean error rate (MER) &Average error rate of FPR and FNR. &avg(FPR,FNR)& 1 & & 
\\\hline

Matthews Correlation Coefficient (MCC) & Classifications quality measure, robust to highly imbalanced data. Here, \textit{x=(TP+FP)(TP+FN)(TN+FP)(TN+FN)} & $\frac{TP\times TN-FP\times FN}{\sqrt{x}}$  & 1 & Suitable for imbalanced data. Unlike precision, recall and f-measure, MCC considers TN & MCC is yet to be widely adopted by the HIDS literature.
\\
\multicolumn{6}{c}{\cellcolor{gray!30}\textbf{Computation Performance (Efficiency)}} \tabularnewline

Time & Training time/ testing time/ execution time & - &	15 & Presents the applicability of the HIDS in a real industry setting in terms of required time.& HIDS efficiency should be used along with effectiveness metrics.
\\\cline{1-5}
\multicolumn{6}{c}{\cellcolor{gray!30}\textbf{Performance of Intermediary Task of Sequence Prediction}} \tabularnewline
BLEU score & Compares a predicted syscall sequence to a target sequence. & brevity penalty$\times$geometric avg of precision score & 3 & Can be used when there is more than one ground truth sequence. Language-independent which makes it straightforward to apply to HIDS models. & Focus on statistical similarities, they can not ensure if the
syscall sequence makes sense to the OS.\\\cline{1-5}
TF-IDF & TF-IDF rises in proportion to the word (i) freq in a document (j), but is offset by the number of documents in the corpus that include the word.& $TF_{i,j}\times log \frac{N}{df_i}$; $TF_{i,j}$=freq of i in j, $N$= doc count, $df_i$= num of doc containing $i$ &2 & Language-independent making it straightforward to apply to HIDS models. &	\\\hline
Cosine Similarity	&  Cosine of the angle between two vectors (x, y) to determine if they are in the same direction.&	$\frac{xy}{||x||||y||}$& 2 & Performs correlation analysis of the predicted and target sequence to ensure if the syscall sequence makes sense to the OS. & Differences in values are not fully considered.\\\hline
\end{tabularx}
\end{table*}
}

\subsubsection{\textbf{Detection performance}} \label{subsubsec:detect_perf} 12 evaluation metrics have been used to validate the detection result by the reviewed HIDS studies which used NLP methods. As shown in Table \ref{tab:metrics}, the mathematical representation of these evaluation metrics includes any of True Positive (TP), False Positive (FP), True Negative (TN) or False Negative (FN) value of confusion matrix \citep{eval_met}. While \textbf{confusion matrix} holds the relevant nominal value instead of normalized values, it is usually used to calculate other metrics (e.g., recall). Detecting all the attacks at a cost of false alarms is highly significant, whereas handling a huge volume of false alarms of a HIDS is time consuming and costly for security analysts. Hence, HIDS usually intends to maximize the attack \textbf{Detection Rate (DR)}  and minimize the \textbf{False Alarm Rate (FAR)}, which are the top 2 most used metrics in the reviewed NLP-based HIDS literature. In the HIDS domain, DR, recall and True Positive Rate (TPR) are referred to as the same. Besides, FAR and False Positive Rate (FPR) are considered as the same \citep{review_ids_5}. 

While optimized \textbf{precision} reveals that all the predicted alarms are worth noticing, \textbf{F-measure} combines both precision and recall \cite{deep_jnca}. F-measure is a better metric for the highly imbalanced HIDS dataset. However, precision, recall and F-measure do not consider TN. Besides, considering the cost of the inability to detect an attack by the HIDS model is high, \textbf{False Negative Rate (FNR)} denotes the missed attacks. Further, a high \textbf{True Negative Rate(TNR)} denotes that the HIDS is usually correct to predict benign behavior. Since the output of the HIDS model is discrete, \textbf{classification error} and \textbf{Mean Error Rate (MER)} are suitable to represent the error of the HIDS model. FNR, TNR, classification error or MER can be used as auxiliary metrics with other metrics (e.g., DR, FAR) \citep{metric}.

The other two highly used metrics in the reviewed NLP-based HIDS literature are \textbf{ROC-AUC} and \textbf{accuracy}. Compared to accuracy, AUC based on the ROC curve (ROC-AUC) provides a more robust measure for the evaluation of HIDS derived from imbalanced datasets as accuracy can be dominated by the majority benign class in HIDS. However, if the dataset is highly imbalanced, the shape of the ROC-AUC curve can be misleading \citep{eval_met}. Thus, Matthews
Correlation Coefficient (MCC) can be used, which is considered one of the best-balanced evaluation measures \citep{eval_met}. Unfortunately, only one study [S4] used MCC. 


\begin{figure*}[]
  \centering
  \includegraphics[width=\textwidth]{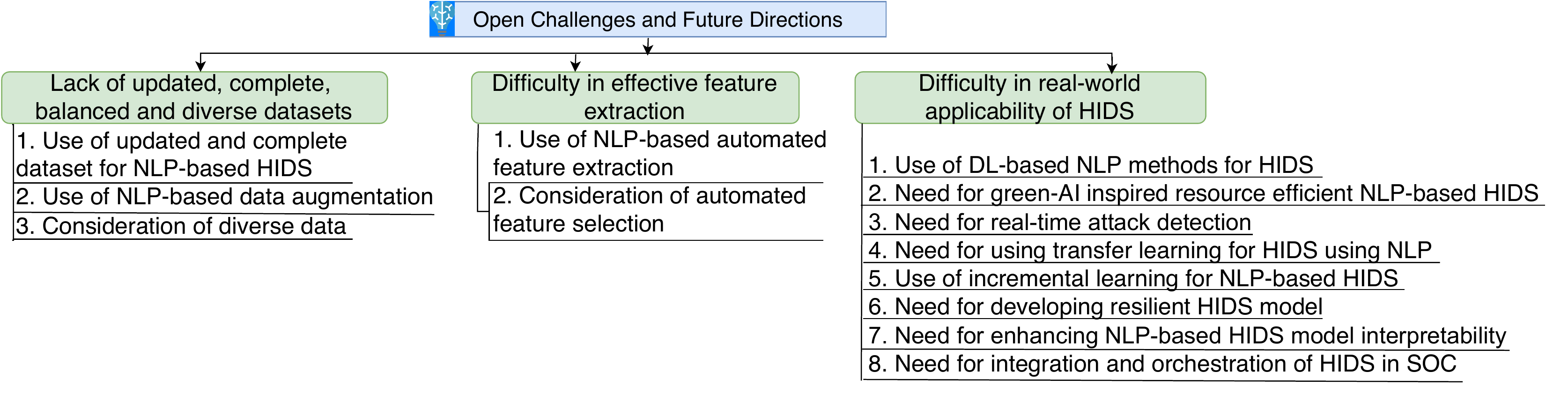}
  \caption{Open Research Challenges and Future Research Directions}
  \label{fig:challenges}
\end{figure*}

\subsubsection{\textbf{Computation performance}} \label{subsubsec:comp_perf}
The computation performance of NLP-based HIDS is measured by time in the reviewed HIDS studies. To evaluate the required time, the reviewed studies reported training time, testing time or execution time. Given the complex structure of DL models (e.g., Seq2Se language modeling), it requires more time and memory even though it provides a high detection accuracy. Unfortunately, only 4 studies [S4, S5, S22, S43] out of 24 studies that use NLP-based DL models have reported time. Furthermore, resource utilization refers to the storage or resource usage (e.g., the size of the stored HIDS model such as learned rules or memory required for training a HIDS model). Evaluation of HIDS in terms of both time and resource utilization is significant to prove the applicability and deployment of the proposed NLP-based HIDS in a real industry setting. Unfortunately, none of the reviewed studies reported the resource utilization of the proposed HIDS using NLP methods. 

\subsubsection{\textbf{Performance of intermediary task of syscall sequence prediction}}\label{subsubsec:intm_perf}
A set of studies performed an intermediary task of syscall sequence prediction before performing the detection. To monitor a system state and predict an attack behavior, syscall sequence prediction is performed. The use of the predicted sequence as supplementary information with the invoked sequence for the detection classifier significantly improved the HIDS performance [S43, S63]. As the performance of HIDS strongly relies on the performance of such intermediary tasks [S44], various metrics have been used to evaluate these intermediary tasks as shown in Table \ref{tab:metrics} with relevant details. 
\textbf{Bilingual Evaluation Understudy Score (BLUE)} [S5, S44, S63] is a benchmark widely utilized in NLP. BLUE Score compares a predicted syscall sequence to a target syscall sequence. The values approaching 1 in the [0, 1] range indicate that the predicted sequence is nearly identical to the target sequence. Another metric is \textbf{Term Frequency-Inverse Document Frequency Score (TD-IDF)} [S44, S63], where the value rises in direct proportion to the frequency of a word in a document, but is offset by the number of documents in the corpus that include the word. Since BLUE and TF-IDF both focus on statistical similarities, they can not ensure if the semantics of the sequence is preserved. Thus, to mitigate this issue, \textbf{Cosine Similarity Score} [S44, S63] is used as well, which performs correlation analysis of the predicted and target sequence.\\ 

\textbf{To summarize RQ2.3}, DR and FAR are the most used metrics in the reviewed NLP-based HIDS literature as HIDS usually intends to maximize the attack DR and minimize the FAR in terms of evaluating detection performance. To justify the applicability of the proposed NLP-based HIDS in real industrial settings, several studies reported the required time in terms of computation performance. However, none of the studies reported the required resource utilization for the NLP-based HIDS. Further, in terms of the performance of intermediary task of syscall sequence prediction, BLUE, TD-IDF and Cosine Similarity are used together to evaluate if the predicted syscall is both syntactically and semantically correct [S44, S63].

\section{Open Research Challenges and Future Directions}\label{subsec:open_issues}
We discuss three critical challenges such as lack of updated, complete, balanced and diverse datasets (Section 7.1), difficulty in effective feature extraction (Section 7.2) and difficulty in real-world applicability (Section 7.3) of NLP-based HIDS. We present 13 potential research directions to address such challenges. Challenges and potential research directions are graphically summarized in Figure \ref{fig:challenges}. We expect Figure \ref{fig:challenges} will benefit practitioners and researchers to get an overview of the open challenges and potential solutions in this domain. 


\subsection{Lack of Updated, Complete, Balanced and Diverse Datasets}
In this section, we present the main issues with the datasets used by the reviewed NLP-based HIDS studies and their potential research directions. 

\subsubsection{\textbf{Use of updated and complete dataset for NLP-based HIDS}} 
The applications, OS and their syscall change rapidly over time with the changing threat landscape, which make outdated datasets less effective for evaluating modern HIDS \citep{dataset, dataset2}. Hence, the HIDS datasets should be kept up-to-date. Yet most of the public datasets are outdated (e.g., the top 3 most used datasets in our review ADFA-LD, UNM and ADFA-WD are almost a decade old). Thus, we recommend to releasing new versions of the existing datasets for handling the concept drift issue [S58] and modern attack scenarios.
    
Furthermore, most of the current IT systems are multi-threaded, still, the available public datasets are not complete as they lack thread information and textual metadata. Thread information and textual metadata are highly important and promising indicators for HIDS \citep{dataset3, dataset4}. For example, ADFA-LD lacks metadata (e.g., parameter, return value) and NGIDS-DS lacks both parameters and more accurate timestamps. Considering only the temporal order of syscall can be susceptible to mimicry attack. This attack can be prevented by combining other system artifacts [S41] (e.g., syscall arguments, function calls and other user-space information). To address these issues, LID-DS dataset \citep{dataset} was developed, which is the first HIDS dataset containing syscall and their timestamps, thread id and diverse metadata of several recent, multi-process and multi-threaded scenarios. Hence, we suggest including the newly available and more complete HIDS datasets for HIDS evaluation rather than using only the decade-old dataset. 

The recommended dataset to be added with the widely adopted ADFA family datasets for evaluation of HIDS is AWSCTD for Windows OS, while NGIDS-DS, LID-DS and PLAID are recommended for Linux OS, as they are more extended and recent datasets that reflects modern host systems and modern attacks.
    

\subsubsection{\textbf{Use of NLP-based data augmentation}} A HIDS prediction model can be biased towards the majority benign class in the commonly used imbalanced datasets. This bias significantly affects the frequently used supervised learning-based HIDS performance. We observe that only 3 out of 25 studies that used supervised learning approach attempted to balance the datasets. These 3 studies proved their effectiveness using diverse methods such as oversampling (SMOTE [S40]), generating malicious samples (GAN [S48], NLP-based SeqGAN and Seq2Seq [S45]). The NLP-based research trend that is gaining popularity from 2019 is various text augmentation approaches (e.g., swap or delete word, and word insertion or replacement based on contextual similar word and word occurrence statistics) \citep{niacin, nlpaug, eda}. A set of studies \citep{sworna2022apiro, easyaug, contextual_data} present the effectiveness of text augmentation in different classifiers, NLP language model and NLP-based prediction model. These augmentation techniques can be investigated to address the data imbalance problem for NLP-based HIDS.


\subsubsection{\textbf{Consideration of diverse datasets}} 
We observe that most studies used only a single dataset (e.g., ADFA-LD (used in 31 studies) and UNM (used in 6 studies)), while only 16 out of 65 studies used multiple datasets for the training and evaluation of HIDS. We recommend using diverse datasets as it will enable the NLP-based HIDS to be platform-independent. It will also improve detection capability and prove the scalability of the NLP-based HIDS in real-life. Firstly, the use of datasets from different OS makes it platform-independent. For example, ADFA-LD, NGIDS-DS are Linux-based and ADFA-WD is from Windows OS. Besides, the use of multiple datasets in the existing studies (e.g., [S44, S49]) would have helped to prove the generalizability of the proposed NLP-based HIDS solution. We recommend NLP-based methods as the NLP-based semantic approach provides high-level portability between diverse OS [S30]. Secondly, the use of large-scale datasets is recommended to represent the real-world scenario to ensure scalability without compromising accuracy. 

\subsection{Difficulty in Effective Feature Extraction}
Training HIDS model using only benign data makes the feature extraction more challenging to find the discriminative features due to the lack of attack data. Besides, manual feature extraction hinders reliable and adaptable feature extraction of HIDS for the changing threat landscape. We discuss the potential research directions to mitigate these challenges.

\subsubsection{\textbf{Use of NLP-based automated feature extraction}} 
The real world adoption of HIDS deals with an enormous amount of continuous data with a rapidly changing threat landscape. So, manually extracted features may become outdated due to feature drift \citep{feature_drift}, and can be easily evaded by attackers. A few studies adopted NLP-based automated feature extraction (e.g., CNN and LSTM-based contextual features) and the use of multi-level feature extraction (e.g., CNN-LSTM) from syscall sequences. However, to extract more effective features for HIDS performance improvement, there is still a demand for taking advantage of recent more sophisticated NLP techniques in HIDS. The advanced NLP techniques such as the use of more feature learning layers using Very Deep Convolutional Neural Networks VDCNN \citep{vdcnn_text}, use of contextual word embedding techniques (e.g., Bidirectional Encoder Representations from Transformers (BERT) \citep{bert}) and deep contextual embedding (e.g., Embeddings from Language Models (ELMo) \citep{elmo}) are yet to be explored in HIDS. Since the use of advanced NLP techniques presents a huge opportunity for feature extraction from continuous data, we recommend that HIDS researchers should focus on reliable and adaptable automated NLP-based feature extraction methods to handle rapidly changing threat landscape.

\subsubsection{\textbf{Consideration of automated feature selection}} 
To reduce dimensionality, computation time and overfitting; while improving ML model interpretability and prediction accuracy, automated feature selection is highly recommended \citep{feature_sel_adv}. Only a few studies explicitly mentioned their feature selection method, where the most used methods (e.g., frequency-based, Principle Component Analysis (PCA)) require manual threshold selection. However, a newer NLP-based approach used in DL models is attention mechanism that can mitigate this limitation, which is adopted by only 4 reviewed studies [S44, S45, S62, S63]. We suggest the use of an NLP-based attention mechanism combined with automated feature extraction to discover more prominent and relevant features from HIDS data.

\subsection{Difficulty in Real-World HIDS Applicability} We propose the following research directions to mitigate the difficulty of ensuring real world applicability of HIDS.

\subsubsection{\textbf{Use of DL-based NLP methods for HIDS}} 
To improve accuracy and reduce FAR for real world applicability of HIDS, further exploration of DL-based NLP methods is recommended. Though TCN is an outstanding alternative to recurrent architecture, it has been used only in 1 study [S8]. Besides, NN with more hidden layers usually provides better performance \citep{vdcnn_text}, so we recommend exploring the deeper versions of CNN, RNN and TCN performance in HIDS. Moreover, the recent practice is to utilize ensemble DL models as it usually outperforms the base models. However, only 3 studies ([S4], [S15], [S46]) used an ensemble of CNN and RNN. Thus, we recommend further focusing on ensembling TCN with other models (e.g., 1D CNN, BiGRU) to explore if it can outperform even deeper versions of CNN, RNN and TCN for HIDS. Moreover, to perform supervised learning using limited annotation data for improved DR and lower FAR, we recommend exploring the effectiveness of using NLP-based low-shot learning or few-shot learning \citep{low_shot_nlp} in HIDS domain.

\subsubsection{\textbf{Need for green-AI inspired resource efficient NLP-based HIDS}}
The adoption of DL model-based NLP method gained popularity in HIDS from 2018 with great improvements in accuracy. However, DL models in NLP are both environmentally unfriendly and prohibitively expensive, which is denoted as ``Red AI” \citep{green_ai}. The required amount of computing resources used to train DL models had increased 300,000-fold from 2012 to 2019 \citep{green_ai}. The necessary computation power of a single DL model in NLP emits 626,000 tonnes $CO_2$, which is five times more than an average car emits throughout its lifetime \citep{green_ai}. Hence, ``Green AI” \citep{green_ai} is encouraged, which considers efficiency as the primary assessment criterion along with effectiveness. Besides, Calero and Piattini \citep{sustainable_design} emphasized that organizations prefer sustainability-based designs (i.e., use of fewer resources for acquiring outcome) in industrial software designs for extensive cost cutting opportunities. Unfortunately, none of the reviewed studies focused on or even reported the required resources (e.g., model size, CPU core hours, disk read and disk write) for the NLP-based HIDS. Hence, we recommend the researchers evaluate and optimize the required resource consumption to ensure the NLP-based HIDS model's applicability as a resource efficient solution in real industrial settings.

\subsubsection{\textbf{Need for real-time attack detection}} 
An organization may fall victim to attacks every 40 seconds \citep{sec_rep}, which demands for real-time attack detection by HIDS. However, only 15 out of 65 studies used the time metric for the NLP-based HIDS evaluation. While the reviewed studies gained high accuracy by applying DL model-based NLP method, they incurred a long training time. For example, a study [S43] using LSTM+VED required 6h 45min and another study [S5] using LSTM and GRU required 50.03min and 43.53min, respectively. To mitigate the requirement of high time issue, we recommend the following methods. Firstly, for syscall sequence data, the use of CNN approaches is more time-efficient as RNN-based (LSTM or GRU) methods encode the input tokens sequentially and operations in RNN-based network structures can not be parallelized, which results in low-efficiency \citep{sentiment_cnn}. Secondly, we recommend using hardware support like GPU to improve the HIDS time efficiency. For example, researchers can explore to improve time efficiency of HIDS based on CNN methods using GPU support \citep{cnn_gpu_1, cnn_gpu2}. Thirdly, HIDS deployed in the industry has to deal with a huge volume of data (e.g., 1 trillion security events are generated by HP per day \citep{bigdata_sec}). Thus, we highly recommend using big data \citep{bigdata} technologies (e.g., Spark, Kafka and Hadoop) for highly scalable HIDS data processing for real-time attack detection.  

\subsubsection{\textbf{Need for using transfer learning for HIDS using NLP}}
Transfer learning is a method that aims to enhance the target domain model's performance by transferring the knowledge which is achieved from a relevant source domain model. Transfer learning reduces the need of a huge amount of target domain data to build a target domain model. Developing individual HIDS with good performance for each computing infrastructure and attack type is difficult, as collecting labeled data and training HIDS from scratch is highly expensive. This issue can be mitigated by transfer learning, which saves training time, resources and improves performance. For example, a study [S12] used transfer learning for improved attack detection performance for a target domain (commonly a domain with little data) utilizing support from a source domain (commonly a domain with huge data). The use of transfer learning by adopting large-scale pre-trained language models \citep{bert, ptm} is a prevalent recent trend in NLP and SE, as it helps to learn universal representation, provides better model initialization for generalization and provides regularization for avoiding over-fitting on a small dataset. Hence, we highly recommend that HIDS researchers explore the creation of diverse pre-trained models and make the models publicly available so that other researchers and developers can adopt these pre-trained models to avail the advantages of transfer learning.



\subsubsection{\textbf{Use of incremental learning for NLP-based HIDS}}  
To keep pace with the rapidly changing threat landscape and to mitigate the concept drift issue, incremental learning-based HIDS is recommended.
Incremental learning is required for the HIDS model to be frequently updated and continuously adapted to new types of attacks and new data with varying characteristics over time. Incremental learning \citep{utgoff1989incremental} uses the previous knowledge and updates the model based on the newly available data reducing the memory requirement and time complexity. Only two of our reviewed studies [S16, S59] adopted incremental learning. In contrast, most of the studies usually use the traditional static batch-retraining method. This method is both time consuming and resource intensive as it discards the previously trained model and trains a new model from scratch on a new dataset including original and new instances. Hence, in accordance with an existing study \citep{bath_retrain}, we emphasize the serious demand for exploring efficient incremental learning algorithms for NLP-based HIDS to address concept drift issues. 

\subsubsection{\textbf{Need for developing resilient HIDS model}} 
The real world deployment of HIDS is susceptible to adversarial attacks. For example, mimicry attack based on syscall temporal order, attacker predicting the alarm generation threshold to keep attack undetected and evasion attacks [S41]. Adversarial attacks strive to evade, undermine or mislead HIDS capabilities. However, only 5 [S30, S35, S41, S42, S59] studies focused on resilient HIDS against adversarial attacks, while 3 studies considered it as future work. A study [S30] showed that the use of NLP-based semantic method inherently makes the HIDS resilient to mimicry attacks. Given the recent prevalence of adversarial attacks targeting models in the cyber security domain, we highly recommend considering the adversarial resilience of the HIDS by adopting methods such as randomness, sanitizing data, adversarial training and using semantic methods \citep{adversarial_ml}. 

\subsubsection{\textbf{Need for enhancing NLP-based HIDS model interpretability}}
HIDS model interpretability raises the transparency of attack predictions, which helps the developers to efficiently debug and refine the model or data for improved performance \citep{chatzimparmpas2020state}. Only a few studies (e.g., [S16, S62]) presented the important features and reasoning of the NLP-based HIDS model's prediction result. HIDS researchers can be motivated by the interpretable NIDS research area \citep{mane2021explaining, wang2020explainable, marino2018adversarial}. Thus, we recommend exploring the applicability of the HIDS interpretability in terms of (1) model-based interpretability (creates a model that is interpretable by nature) and (2) post hoc (applies an interpretability approach after training a black box model) \citep{murdoch2019definitions}. The NLP approaches such as locating significant n-grams in sentences based on the intermediate outputs of CNN can provide interpretability \citep{automating_intention} that can be adopted in NLP-based HIDS.

\subsubsection{\textbf{Need for integration and orchestration of HIDS in SOC}} Security Operation Center (SOC) of an organization uses 76 security tools on average \citep{sec_tools} including HIDS. Even the use of multiple HIDS to protect different hosts is a common practice in SOC \citep{select_ids}. Hence, HIDS should be able to collect and process data from diverse tools. For example, MISP (Malware Information sharing Platform) \citep{MISP_official} shares cyber security indicators to be integrated with HIDS. Further, the never-ending flow of alerts generated by HIDS tools need to be correlated by Security and Information Management (SIEM) system or Security Orchestration, Automation and Response (SOAR) Platform. For example, a SIEM tool named Splunk \citep{splunk_siem} enables to search, analyze and visualize the alerts gathered from tools (e.g., HIDS) of the SOC. Hence, data interpretability and interoperability of HIDS tools is a key requirement to ensure the integration and orchestration \citep{sec_orch_caise} of HIDS in a SOC, which has not been focused on in the reviewed studies. Recently, only 3 studies [S13, S26, S33] utilized semantic ontology for intrusion detection by integrating IDS/IPS sensor information (e.g., HIDS, NIDS), sensor data streams, malware data (Symantec's website), web text data, and domain expert knowledge. To mitigate the gap between academia and industry, we recommend that HIDS researchers should focus on the existing tools and techniques available in the industry and propose methods to integrate them with the proposed HIDS in SOC scenario. This leads to the following future research directions. Firstly, we recommend comparative evaluation of proposed HIDS with existing HIDS tools (e.g., open-source WAZUH or OSSEC \citep{ossec}). Secondly, we recommend incorporating the available updated threat intelligence data sources to HIDS for attack detection and ensuring the HIDS alerts/output to be interpretable by the SIEM or SOAR tools for smooth integration and orchestration of proposed HIDS in real SOC scenarios. Further, we suggest that semantic ontology and reasoning can be explored to integrate and orchestrate \citep{sec_orch_caise} HIDS with other security tools, which will validate the use of the proposed HIDS in the real SOC scenario. 


\section{Threats to Validity}\label{sec:threats}

We carefully followed the widely adopted SLR guidelines \citep{slr_guidelines} to design and conduct our SLR. We adopted suitable steps to mitigate the effects of identified threats to validity of this SLR as presented below:

\textit{Search Strategy}: Missing some relevant studies is a common threat to an SLR. To minimize this effect, we used Scopus (the most comprehensive search engine with the largest indexing system \citep{roland}) and complemented it with the two most frequently used digital libraries, IEEE Xplore and ACM Digital Library \citep{digital_lib}. Moreover, we ran a series of pilot searches to find a suitable search string to ensure the retrieval of the relevant papers. Besides, both forward and backward snowballing techniques were conducted to find other relevant papers that are overlooked by the search string.

\textit{Selection process}: The study selection process may be affected by the authors’ subjective judgment. To mitigate this threat, we performed a multi-step process (Section \ref{subsec:study_selection}) with clearly specified inclusion-exclusion criteria to select the relevant studies. We also defined specific quality assessment criteria to exclude low-quality papers. At each step of the selection process, we discussed and resolved ambiguities to minimize the selection bias.
    
\textit{Data Extraction and Synthesize}: Results and findings may be influenced by human error and author bias in data extraction, data analysis and data interpretation. To address this issue, a data extraction form was created and iteratively improved to collect sufficient and consistent information required to answer the RQs. Besides, all the data-extraction activities, data synthesizing and interpretation of our quantitative and qualitative analysis were cross-checked by the authors. All the disagreements were discussed and resolved through discussions. 


\section{Conclusion} \label{sec: concl}
This paper presents an SLR aimed at systematically and rigorously selecting and analyzing the existing literature on NLP-based HIDS. The findings are expected to form an evidence-based body of knowledge of taxonomic analysis of the NLP methods, attacks, datasets and evaluation metrics used in NLP-based HIDS for practical use in the industrial setting. We synthesized 65 papers from the last decade on NLP-based HIDS. We categorized and compared the NLP-based HIDS solutions, datasets and evaluation metrics to help developers select a suitable type for a specific HIDS application. We discussed the role of NLP in HIDS and the impact of attacks that are detected by NLP-based HIDS.

Our review aims to help researchers by providing an overview of this burgeoning research landscape. The increasing number of studies in NLP-based HIDS shows the significantly growing attention among the research community as 42 papers were published in the last 4 and a half years. Yet our review identified crucial open issues and proposed a roadmap of future work to help the researchers to mitigate those issues. The implications for the researchers are as follows:
\begin{enumerate}[noitemsep,topsep=0pt]

\item We recommend the researchers to focus on real-time accurate HIDS development leveraging the latest NLP techniques (e.g., text augmentation, NLP-based low-shot learning). Besides, researchers should focus on adopting incremental learning, enhancing HIDS model interpretability with robustness to adversarial attacks using NLP. Moreover, researchers should focus on semantic integration of HIDS in SOC. 
\item We encourage the researchers to adopt NLP-based transfer learning using pre-trained HIDS models as it helps to learn universal representation, provides better model initialization for generalization and provides regularization for avoiding over-fitting on small HIDS dataset.
\item	Our findings advocate the demand for further research in DL-based NLP methods for developing HIDS as the recent practice of adopting DL-based NLP approaches showed significantly better performance compared to the prevalently used traditional ML and rule based approaches. 
\item	We recommend the researchers to get inspired by green-AI and focus on time and resource efficiency of the proposed NLP-based HIDS along with effectiveness to provide cost cutting opportunities for real industrial adoption. 
\item	We encourage the researchers to make their dataset publicly available as 8 studies used private datasets, where the inaccessibility to the dataset hinders research advancement.  
\item	We recommend researchers to use MCC, which is a suitable metric for the most used HIDS imbalanced datasets. 
\item	Our comprehensive taxonomy is expected to help researchers to frame future research in this domain.  
\end{enumerate}

Our findings are expected to help practitioners potentially utilize the NLP methods as we highlight the existing prevalent practices and considerations in NLP-based HIDS. The implications for the practitioners are as follows:
\begin{enumerate}[noitemsep,topsep=0pt]
\item Practitioners can make well-informed decisions while developing a HIDS based on our critically reviewed end-to-end pipeline of NLP-based HIDS. We identified, categorized and highlighted the strengths and weaknesses of each of the categories for the used NLP methods, key factors to consider while developing HIDS, datasets and evaluation metrics. 

\item	For securing critical infrastructure, we suggest the practitioners adopt the big data frameworks and sophisticated hardware support such as GPU for real-time detection by HIDS. 
\item	Our findings guide the practitioners by providing the prevalent practices such as the dominant use of semi-supervised learning approach due to the lack of balanced datasets and the use of anomaly-based approach to detect unknown attacks (e.g., zero-day attack). We recommend that practitioners use hybrid approaches to gain the benefit of both signature and anomaly-based detection approaches. 
\item	We recommend the practitioners to train and validate the HIDS model with their industry-specific data before the deployment of the target HIDS, as our review found that most studies used public datasets, which are mostly outdated and lack sufficient and diverse attack instances.
\item	We encourage the practitioners to analyze the trade-off between detection performance and computation performance (i.e., required time and resource) while choosing a HIDS model to achieve maximum detection rate with a lower false alarm rate at minimum processing time and cost. Unfortunately, most of the studies did not report time and resource utilization, which questions the applicability of the proposed NLP-based HIDS in the industrial setting. Besides, we encourage the practitioners to frequently share their experiences and the recent attacks’ details encountered in real industrial settings, which can help to mitigate the gap between academic research and industrial demand. Practitioners sharing the attack signature will help to store the up-to-date signature for signature-based detection and will help to evaluate the anomaly-based models' capability to detect new attacks.
\end{enumerate}

\section*{Acknowledgement}
The work has been supported by the Cyber Security Research Centre Limited whose activities are partially funded by the Australian Government’s Cooperative Research Centres Programme.
%


\section{Appendix}

\subsection{List of Selected Papers} \label{subsec:paper_list}
{\small
\begin{itemize}  
    \item [S1] Z. Liu, N. Japkowicz, R. Wang, Y. Cai, D. Tang, \& X. Cai, ``A statistical pattern based feature extraction method on system call traces for anomaly detection", Information and Software Technology, 2020.
    \item [S2] Y. Shin, \& K. Kim, ``Comparison of anomaly detection accuracy of host-based intrusion detection systems based on different machine learning algorithms", International Journal of Advanced Computer Science and Applications, 2020.
    \item [S3] X. Zhang, Q. Niyaz, F. Jahan, \& W. Sun, ``Early Detection of Host-based Intrusions in Linux Environment", IEEE International Conference on Electro Information Technology, 2020.
    \item [S4] D. Čeponis, \& N. Goranin, ``Investigation of dual-flow deep learning models LSTM-FCN and GRU-FCN efficiency against single-flow CNN models for the host-based intrusion and malware detection task on univariate times series data", Applied Sciences, 2020.
    \item [S5] L. Bouzar-Benlabiod, S. H. Rubin, K. Belaidi, \& N. E. Haddar, ``RNN-VED for Reducing False Positive Alerts in Host-based Anomaly Detection Systems", IEEE 21st International Conference on Information Reuse and Integration for Data Science, 2020.
    \item [S6] B. Subba, \& P. Gupta, ``A tfidfvectorizer and singular value decomposition based host intrusion detection system framework for detecting anomalous system processes", Computers \& Security, 2021.
    \item [S7] X. Liao, C. Wang, \& W. Chen, ``Anomaly Detection of System Call Sequence Based on Dynamic Features and Relaxed-SVM", Security and Communication Networks Journal, 2022
    \item [S8] N. Fu, N. Kamili, Y. Huang, \& J. Shi, ``A Novel Deep Intrusion Detection Model Based On a Convolutional Neural Network", . Aust. J. Intell. Inf. Process. Syst., 2019.
    \item [S9] M. Grimmer, T. Kaelble, \& E. Rahm, ``Improving Host-Based Intrusion Detection Using Thread Information", Communications in Computer and Information Science, 2021.
    \item [S10] D. Čeponis, \& N. Goranin, ``Evaluation of deep learning methods efficiency for malicious and benign system calls classification on the AWSCTD", Security and Communication Networks, 2019.
    \item [S11] S. Suratkar, F. Kazi, R. Gaikwad, A. Shete, R. Kabra, \& S. Khirsagar, ``Multi Hidden Markov Models for Improved Anomaly Detection Using System Call Analysis", IEEE Bombay Section Signature Conference, 2019.
    \item [S12] O. Ajayi, \& A. Gangopadhyay, ``DAHID: Domain adaptive host-based intrusion detection", IEEE International Conference on Cyber Security and Resilience, 2021.
    \item [S13] Ö. Can, M. O. Ünallır, E. Sezer, O. Bursa, \& B. Erdoğdu, ``A semantic web enabled host intrusion detection system", International Journal of Metadata, Semantics and Ontologies, 2018.
    \item [S14] Y. Lu, \& S. Teng., ``Application of Sequence Embedding in Host-based Intrusion Detection System", IEEE 24th International Conference on Computer Supported Cooperative Work in Design, 2021
    \item [S15] A. Chawla, B. Lee, S. Fallon, \& P. Jacob, ``Host based intrusion detection system with combined CNN/RNN model", Joint European Conference on Machine Learning \& Knowledge Discovery in Databases, 2018.
    \item [S16] P. F. Marteau, ``Sequence covering for efficient host-based intrusion detection", IEEE Transactions on Information Forensics and Security, 2018.
    \item [S17] N. N. Tran, R. Sarker, \& J. Hu, ``An approach for host-based intrusion detection system design using convolutional neural network", International conference on mobile networks and management, 2017.
    \item [S18] F. J. Mora-Gimeno, H. Mora-Mora,  B. Volckaert, \& A. Atrey, ``Intrusion Detection System Based on Integrated System Calls Graph and Neural Networks". IEEE Access, 2021.
    \item [S19] B. Subba, S. Biswas, \& S. Karmakar, ``Host based intrusion detection system using frequency analysis of n-gram terms", TENCON IEEE Region 10 Conference, 2017.
    \item [S20] Z. Hu, L. Liu, H. Yu, \& X. Yu, ``Using Graph Representation in Host-Based Intrusion Detection", Security and Communication Networks, 2021.
    \item [S21] J.H. Ring, C.M. Van Oort, S. Durst, V. White, J.P. Near, \& C. Skalka, ``Methods for Host-based Intrusion Detection with Deep Learning", Digital Threats: Research and Practice, 2021.
    \item [S22] Y. Zhang, S. Luo, L. Pan, \& H. Zhang, ``Syscall-BSEM: Behavioral semantics enhancement method of system call sequence for high accurate and robust host intrusion detection", Future Generation Computer Systems, 2021.
    \item [S23] S. S. Murtaza, W. Khreich, A. Hamou-Lhadj,  \& S. Gagnon, ``A trace abstraction approach for host-based anomaly detection", IEEE Symposium on Computational Intelligence for Security and Defense Applications, 2015.
    \item [S24] S. S. Murtaza, A. Hamou-Lhadj, W. Khreich, \& M. Couture, ``Total ADS: Automated software anomaly detection system", IEEE 14th International Working Conference on Source Code Analysis and Manipulation, 2014.
    \item [S25] M. B. L. M. Anandapriya, \& B. Lakshmanan, ``Anomaly based host intrusion detection system using semantic based system call patterns", IEEE 9th International Conference on Intelligent Systems \& Control, 2015.
    \item [S26] S. N. Narayanan, A. Ganesan, K. Joshi, T. Oates, A. Joshi, \& T. Finin, ``Early detection of cybersecurity threats using collaborative cognition", IEEE 4th international conference on collaboration and internet computing, 2018.
    \item [S27] K. H. Cha, \& D.Ki. Kang, ``Empirical analysis of effective misuse intrusion detection by trace classification using conditional random fields", Journal of Engineering and Applied Sciences, 2017.
    \item [S28] M. Xie, J. Hu, X. Yu, \& E. Chang, ``Evaluating host-based anomaly detection systems: Application of the frequency-based algorithms to ADFA-LD", International Conference on Network \& System Security, 2014.
    \item [S29] M. Xie, J. Hu, \& J. Slay, ``Evaluating host-based anomaly detection systems: Application of the one-class SVM algorithm to ADFA-LD", 11th International Conference on Fuzzy Systems \& Knowledge Discovery, 2014.
    \item [S30] G. Creech, \& J. Hu, ``A semantic approach to host-based intrusion detection systems using contiguousand discontiguous system call patterns", IEEE Transactions on Computers, 2013.
    \item [S31] M. Xie, \& J. Hu, ``Evaluating host-based anomaly detection systems: A preliminary analysis of adfa-ld", 6th International Congress on Image and Signal Processing, 2013.
    \item [S32] A. Mustafa, M. Solaimani, L. Khan, K. Chiang, \& J. Ingram, ``Host-based anomaly detection using learning techniques", IEEE 13th International Conference on Data Mining Workshops, 2013.
    \item [S33] S. More, M. Matthews, A. Joshi, \& T. Finin, ``A knowledge-based approach to intrusion detection modeling", IEEE Symposium on Security and Privacy Workshops, 2012.
    \item [S34] A. Sultana, A. Hamou-Lhadj, \& M. Couture, ``An improved hidden markov model for anomaly detection using frequent common patterns", IEEE International Conference on Communications, 2012.
    \item [S35] W. Khreich, S. S. Murtaza, A. Hamou-Lhadj, \& C. Talhi, ``Combining heterogeneous anomaly detectors for improved software security", Journal of Systems and Software, 2018.
    \item [S36] N. A. Milea, S. C. Khoo, D. Lo,  \& C. Pop, ``Nort: Runtime anomaly-based monitoring of malicious behavior for windows", International Conference on Runtime Verification, 2011.
    \item [S37] G. Serpen, \& E. Aghaei, ``Host-based misuse intrusion detection using PCA feature extraction and kNN classification algorithms", Intelligent Data Analysis, 2018.
    \item [S38] T. H. Lee, H. Y. Huang, \& C. Juang, ``A high-performance deep learning architecture for host-based intrusion detection system", IEEE REGION 10 CONFERENCE, 2020.
    \item [S39] W. Wang, Z. Yang, \& M. Zhang, ``Intrusion Detection Technology Based on Rough Set Attribute Reduction Theory", International Conference on Human Centered Computing, 2016.
    \item [S40] E. Aghaei, \& G. Serpen, ``Ensemble classifier for misuse detection using N-gram feature vectors through operating system call traces", International Journal of Hybrid Intelligent Systems, 2017.
    \item [S41] W. Khreich, B. Khosravifar, A. Hamou-Lhadj, \& C. Talhi, ``An anomaly detection system based on variable N-gram features and one-class SVM", Information and Software Technology, 2017.
    \item [S42] B. Borisaniya, \& D. Patel, ``Evaluation of modified vector space representation using adfa-ld and adfa-wd datasets", Journal of Information Security, 2015.
    \item [S43] L. Bouzar-Benlabiod, L. Méziani, S. H. Rubin, K. Belaidi, \& N. E. Haddar, ``Variational encoder-decoder recurrent neural network (VED-RNN) for anomaly prediction in a host environment", IEEE 20th International Conference on Information Reuse and Integration for Data Science, 2019.
    \item [S44] S. Lv, J. Wang, Y. Yang, \& J. Liu, ``Intrusion prediction with system-call sequence-to-sequence model", IEEE Access, 2018.
    \item [S45] S. Shin, I. Lee, \& C. Choi, ``Anomaly dataset augmentation using the sequence generative models", 18th IEEE International Conference On Machine Learning And Applications, 2019.
    \item [S46] N. N. Diep, N. T. T. Thuy, \& P. H. Duy, ``Combination of multi-channel CNN and BiLSTM for host-based intrusion detection", Southeast Asian Journal of Sciences, 2018.
    \item [S47] S. Wunderlich, M. Ring, D. Landes, \& A. Hotho, "Comparison of system call representations for intrusion detection", International Joint Conference: 12th International Conference on Computational Intelligence in Security for Information Systems and 10th International Conference on European Transnational Education, 2019.
    \item [S48] K. Kim, ``GAN based Augmentation for Improving Anomaly Detection Accuracy in Host-based Intrusion Detection Systems", International Journal of Engineering Research and Technology, 2020.
    \item [S49] S. Jose, D. Malathi, B. Reddy, \& D. Jayaseeli, ``Anomaly based host intrusion detection system using analysis of system calls", International Journal of Pure \& Applied Mathematics, 2018.
    \item [S50] A. Chawla, P. Jacob, B. Lee, \& S. Fallon, ``Bidirectional LSTM autoencoder for sequence based anomaly detection in cyber security", International Journal of Simulation--Systems, Science \& Technology, 2019.
    \item [S51] S. Wunderlich, M. Ring, D. Landes, \& A. Hotho, ``The Impact of Different System Call Representations on Intrusion Detection", Logic Journal of the IGPL, 2020.
    \item [S52] Q. Quan, W. Jinlin, Z. Wei, \& X. Mingjun, ``Improved edit distance method for system call anomaly detection", IEEE 12th International Conference on Computer and Information Technology, 2012.
    \item [S53] N. Hubballi, S. Biswas, \& S. Nandi, ``Sequencegram: n-gram modeling of system calls for program based anomaly detection", 3rd International Conference on Communication Systems \& Networks, 2011.
    \item [S54] R. Vinayakumar, M. Alazab, K. P. Soman, P. Poornachandran, A. Al-Nemrat, \& S. Venkatraman, ``Deep learning approach for intelligent intrusion detection system", IEEE Access, 2019.
    \item [S55] Z. Wang, Y. Liu, D. He, \& S. Chan, ``Intrusion detection methods based on integrated deep learning model", computers \& security, 2021.
    \item [S56] N. Hubballi, ``Pairgram: Modeling frequency information of lookahead pairs for system call based anomaly detection", 4th International Conference on Communication Systems and Networks, 2012.
    \item [S57] J. Liu, K. Xiao, L. Luo, Y. Li, \& L. Chen, ``An intrusion detection system integrating network-level intrusion detection and host-level intrusion detection", IEEE 20th International Conference on Software Quality, Reliability and Security, 2020.
    \item [S58] L. Cheng, Y. Wang, Y. Zhou, \& X. Ma, ``EADetection: An efficient and accurate sequential behavior anomaly detection approach over data streams", International Journal of Distributed Sensor Networks, 2018.
    \item [S59] Y. S. Ikram, \& M. A. Madkour, ``Enhanced Host-Based Intrusion Detection Using System Call Traces", Journal of King Abdulaziz University-Computing and Information Technology Sciences, 2019.
    \item [S60] W. Liu, L. Ci, \& L. Liu, ``A new method of fuzzy support vector machine algorithm for intrusion detection", Applied Sciences, 2020.
    \item [S61] E. N. Yolacan, J. G. Dy, \& D. R. Kaeli, ``System call anomaly detection using multi-hmms", IEEE 8th International Conference on Software Security and Reliability-Companion, 2014.
    \item [S62] Z. Q. Qin, X. K. Ma, \& Y. J. Wang, ``ADSAD: An unsupervised attention-based discrete sequence anomaly detection framework for network security analysis", Computers \& Security, 2020.
    \item [S63] G. Sarraf, \& M. S. Swetha, ``Intrusion Prediction and Detection with Deep Sequence Modeling", In International Symposium on Security in Computing and Communication, 2019.
    \item [S64] M. S. Islam, K. K. Sabor, A. Trabelsi, W. Hamou-Lhadj, \& L. Alawneh, ``MASKED: A MapReduce Solution for the Kappa-pruned Ensemble-based Anomaly Detection System", IEEE International Conference on Software Quality, Reliability and Security, 2018.
    \item [S65] M. Raj, \& S. Jose, ``A Host Based Intrusion Detection System Using Improved Extreme Learning Machine", International Journal for Innovative Research in Science and Technology, 2015.
\end{itemize}
}
\subsection{Table of Notation}\label{subsec:notation}
Table 13 shows the frequently used acronyms' abbreviations.

\begin{table}[h]
\centering
\caption{Table of Notation}
\label{tab:abbr}
\footnotesize
\begin{tabularx}{\columnwidth}{lX|lX}
\toprule
Acronym & Abbreviation & Acronym & Abbreviation\\
\midrule
DL & Deep Learning  & ML & Machine Learning \\
SLR & Systematic Literature Review  & NLP & Natural Language Processing \\ 
HIDS & Host Intrusion Detection System & SOC & Security Operation Center\\
Syscall & System Call   & FAR & False Alarm Rate\\\hline
\end{tabularx}
\end{table}

\appendix

\bibliographystyle{elsarticle-harv} 
\bibliography{elsarticle-template-harv.bib}





\end{document}